\input harvmac
\input epsf
\baselineskip=.55truecm
\Title{\vbox{\hbox{HUTP--95/A048, BRX-TH-388}\hbox{hep-th/9601023}}}
{\vbox{\centerline{A Critical Behaviour of Anomalous Currents,}
\centerline{Electric-Magnetic Universality and CFT$_4$}}}
\vskip .1in
\centerline{\sl D. Anselmi$\,^{a,}$\foot{E-mail:
anselmi@string.harvard.edu},
M. Grisaru$\,^{b,}$\foot{E-mail: grisaru@binah.cc.brandeis.edu} and
A. Johansen$\,^{a,}$\foot{E-mail: johansen@string.harvard.edu}}
\vskip .2in
\centerline{$^a$\it Lyman Laboratory of Physics, Harvard University}
\centerline{\it Cambridge, MA 02138, USA}
\centerline{$^b$\it Physics Department, Brandeis University}
\centerline{\it Waltham, MA 02254, USA}
\vskip .2in
\centerline{ABSTRACT}
\vskip .2in

We discuss several aspects of  superconformal field theories
in four dimensions (CFT$_4$), in the context of electric-magnetic
duality.
We analyse the behaviour of anomalous currents
under RG flow to a conformal fixed point in $N=1$, $D=4$
supersymmetric gauge theories.
We prove that the anomalous dimension of the Konishi
current  is related to the slope of the beta
function at the critical point.
We extend the duality map to the (nonchiral) Konishi current.
As a byproduct we compute the slope of the beta function in the
strong coupling
regime. We note that
 the OPE of $T_{\mu\nu}$ with itself does not close, but mixes with
a  special additional operator $\Sigma$ which  in general
  is the Konishi current. We discuss the implications of this fact  
in generic
interacting conformal theories.  In particular, a SCFT$_4$ seems to be
naturally
equipped with a privileged off-critical deformation
$\Sigma$ and this allows us to argue
that electric-magnetic duality can be extended to a neighborhood
of the critical point.
We also stress that in  SCFT$_4$
there are {\sl two} central charges, $c$ and $c'$,
associated with the stress tensor and  $\Sigma$, respectively;
$c$ and $c'$ allow us to count both the vector multiplet and the matter
multiplet
effective degrees of freedom of the theory.

\Date{\bf \phantom{December 1995}}

\newsec{Introduction}
A recent insight into quantum field theory has  been
the discovery by Seiberg of the electric-magnetic duality of
non-trivial N=1
superconformal
theories in four dimensions that can be realized as infrared fixed
points of
ordinary N=1 supersymmetric theories \ref\emd{N. Seiberg, Nucl. Phys.
{\bf B 435} (1995) 129.}.
Following his approach other examples of such
 dynamics have been found \ref\exam{See for a review
K. Intriligator and N. Seiberg, {\it Lectures on Supersymmetric Gauge
Theories and Electro-Magnetic Duality},
 Nucl. Phys. Proc. Suppl.  {\bf 45BC} (1996) 1; hep-th/9509066.}.
All these examples have survived various  self-consistency tests
and therefore one may be convinced that this understanding of their
renormalization group flow is correct.
However the physical information about  the resulting
superconformal theories
seems to be incomplete.
In particular  we do not know the exact spectrum and physical
correlators of the
theory
at the conformal point
\ref\HW{An analysis of the physical correlators
in
conformal theories with extended supersymmetry has been undertaken in
P. Howe and P. West, {\it Non-Perturbative Green's Functions in
Theories with Extended Superconformal Symmetry}.
Preprint KCL-TH-95-9; hep-th/9509140.}. We also have only limited  
information
about duality maps between operators of these theories.
A non-trivial duality map of primary chiral
operators at the
critical point
has been constructed in many models \exam , \ref\kutasov{D. Kutasov,
Phys. Lett. {\bf 351B} (1995) 230, hep-th/9503086\semi
D. Kutasov and A. Schwimmer, 
Phys. Lett. {\bf 354B} (1995) 315, hep-th/9505004\semi
D. Kutasov, N. Seiberg and A. Schwimmer, 
{\it Chiral Rings, Singularity Theory
and Electric-Magnetic Duality}.  Nucl. Phys. {\bf B459} (1996) 455;
hep-th/9510222.} but the extension to nonchiral operators is more  
problematic
\foot{For a recent discussion of some aspects of
such an extension see ref.
\ref\berkooz{M. Berkooz, {\it A comment on Non-Chiral Operators in SQCD
and its Dual}. Nucl. Phys. {\bf B466} (1996) 75; hep-th/9512024.}.}.

In a recent paper \ref\ksv{I. Kogan, M. Shifman and
A. Vainshtein, {\it Matching Conditions and
Duality in N=1 SUSY Gauge Theories in the Conformal Window}. 
Phys. Rev. {\bf D53} (1996) 4526;  hep-th/9507170.}
an analysis of the infrared behaviour of the R-current has been presented.
In this paper, we discuss some other aspects of critical  
behaviour of N=1
supersymmetric theories, focusing on the role and properties of anomalous
currents.
We shall be mostly interested in the Konishi axial current,  
contained in a
superfield that we call the Konishi superfield: $J  \sim  
\bar{\Phi}\Phi$  where
$\Phi$ denotes the chiral superfields of these theories.
Away from criticality  the Konishi current  is anomalous - the anomaly is
proportional to the gauge $\beta$-function  - and is
not considered in the analysis of the
infrared dynamics since it does not generate any symmetry. One may  
ask what is
the fate of this anomalous  current at the
critical point.
A possible answer would be that  since the anomaly
is proportional to the  beta function, the current  becomes  
conserved since the
beta function vanishes.
This  would be similar to the fate of the trace anomaly,
which is nonvanishing
away from criticality and proportional to the beta function,
but disappears at the critical point.

In the case of the Konishi current, if the above scenario were  
realised the
conformal theory would have  additional global
symmetries, generated by the newly conserved current,
as compared
to the non-critical theory.
Such a scenario would  also
imply that there is a jump in the cohomology
ring of the BRST operator in the twisted N=1 supersymmetric theory
\ref\joh{A. Johansen, Mod. Phys. Lett. {\bf A9} (1994) 2611;
Nucl. Phys. {\bf B 436} (1995) 291.}.
(Another possibility is that the anomalous current decouples from  
the physical
correlators, implying that the related degrees of freedom
become massive near the conformal point.)

In the present paper we demonstrate that  although the anomaly for the {\it
renormalized}
Konishi current is
proportional to the
beta function, this anomalous current survives at the conformal
fixed point and  remains anomalous. Moreover, it plays an important  
role in
quantum conformal
field theory in four dimensions (CFT$_4$) since it appears in the  
operator
product expansions of the stress energy tensor,
and defines a new central charge.

As we shall show, at large distances, the two-point correlator of   
two {\it
renormalized}
Konishi superfields behaves like $<J(x) J(y)>\sim
[\beta(g^2)/g^4]^2 |x-y|^{-4}$ where $g= g(|x-y|)$ is the
running gauge coupling constant
at the scale $1/|x-y|$ and
$\beta$ is the Gell-Mann-Low function.
Consequently, in a theory with a non-trivial conformal fixed point
this correlator  has a power-like behaviour $\sim 1/|x-y|^{4+2\delta}$
for  $|x-y|\to \infty$, where $\delta$, the anomalous dimension of  
the Konishi
current, is  given by the slope of the beta function at criticality:
\eqn\slope{\delta =\beta' (\alpha_*)}
Here $\alpha = g^2/4 \pi$.
To understand this relation we note  that the Konishi current
is a relevant deformation of the critical theory (due to the  
Konishi anomaly,
adding the Konishi operator to the action is  equivalent to  
changing the gauge
coupling constant), so that
Eq. \slope \ can be seen as the generalization of the
analogous property that holds in two dimensions
\ref\zam{A. Zamolodchikov, Yad. Fiz. {\bf 46} (1987) 1819.}.
In particular in
SUSY QCD with the $SU(N_c)$ gauge group and $N_f$ flavours of
matter superfields in fundamental representation
($3N_c/2\leq N_f\leq 3N_c$) without superpotential,
we find that for  the case $\epsilon=3-N_f/N_c<<1$
(which corresponds to a perturbative fixed point \emd )
the  anomalous dimension
of the Konishi current  is $\delta =\epsilon^2/3$ at the conformal point.

The above result  can be generalized to  models with superpotential.
We use such a generalization to construct a duality map for  the  
(non-chiral)
Konishi operator of the electric theory.
Using Seiberg electric-magnetic duality \emd\ one can argue that the
conformal dimension of the electric Konishi current is equal to  
that of its
magnetic counterpart in the dual formulation.
Consequently we find a relation between the slopes of
the beta functions in these two formulations, in the conformal window.
In particular we compute the dimension
of the Konishi current at the strongly interacting conformal point
$\sigma =-3/2+N_f/N_c<<1$ in the  electric theory without superpotential,
which is
small in this regime. We find  $ \delta = \beta'  
(\alpha_*)=28\sigma^2/3$.

Eq.\slope\  is analogous to the fact that the operator of the trace  
anomaly
$\Theta =\beta (\alpha) F_{\mu\nu}F^{\mu\nu}$
acquires the anomalous dimension $\beta ' (\alpha)>0$
(for a recent discussion see ref. \ref\sonoda{H. Sonoda,
{\it The Energy-Momentum Tensor in Field Theory} I. hep-th/9504133.}).
Despite the fact that the operator $\Theta$
survives in the infrared the energy-momentum tensor is traceless
at the conformal fixed point because the traceless part has
canonical dimension which is smaller than that of $\Theta$.
The story is different for the Konishi supermultiplet $J$ all  
components of
which
have the same anomalous dimension at the fixed point.
In particular,  the  Konishi axial current $a_{\mu}$  has  a  
non-canonical
dimension there.This is compatible with the fact that $a_{\mu}$
does not generate any new symmetry of the theory  at the conformal  
point; nor
is there a jump in the cohomology
ring of  the BRST operator  in the twisted N=1 supersymmetric theory.
Furthermore, the  (component) anomaly operators  which enter the chiral
superfield
$\bar{D}^2 J$ have  conformal dimensions which do not satisfy
the relation $d=3R/2$ between $R$ charge and
conformal dimension $d$ which follows from the $N=1$ superconformal
algebra
\ref\sca{See, for example, E.S. Fradkin and  A.A. Tseytlin,
Phys. Rep. {\bf 119}  (1985) 233.}.
The explanation of this phenomenon is that
$\bar{D}^2 J$ is not a primary chiral operator of the superconformal
algebra at the fixed point.

The conserved currents  contained in the supercurrent superfield
$J_{\alpha\dot{\alpha}}$ have of course their canonical dimensions.
However, away from the fixed point,  the anomalous axial current  
that enters
the supercurrent superfield
$J_{\alpha\dot{\alpha}}$ is   a linear combination of
a conserved $R$ current and the Konishi current.
Therefore it contains  parts with different scaling behaviour.
This is consistent with  superPoincar\'e
symmetry due to the reducibility of the superfield  
$J_{\alpha\dot{\alpha}}$
at criticality. Furthermore,
it follows from the fact that the conformal dimension of the
Konishi current
is larger then its canonical dimension, that the Seiberg non-anomalous
$R$ current
coincides with the current which enters $J_{\alpha\dot{\alpha}}$ in the
infrared.
This last fact was first established in ref. \ksv\
by analysing operator equations for anomalous currents.

The importance of anomalous currents at criticality manifests itself also
in  the nonclosure
of the operator product expansion of two  stress tensors or supercurrent
superfields: indeed an additional
operator $\Sigma$ appears that is not conserved
and generally coincides  with the Konishi superfield
(apart from a multiplicative factor that can depend on the coupling).
One possible implication of this is an extension of the
idea of electric-magnetic duality. Indeed, mapping two conformal theories
amounts to identifying their  superfields $J_{\alpha\dot\alpha}$.
This, in turn, implies the identification of their OPE's and, in  
particular,
of the operator $\Sigma$. Consequently, one has a natural mapping between
two
deformations of the theories, at least in neighborhoods of the  
critical points.

As mentioned above, away 
 from the fixed point the Konishi current is the only renormalizable
deformation
of conformal SQCD. With this interpretation we
find that the subleading behaviour of the electric and magnetic theories
near the conformal fixed point shows an unexpected universality.
This universality is a consequence  of electric-magnetic duality
and means
that
in the infinitely small neighborhood of a critical point there
exists a map from
electric off-critical theory to the magnetic one.  We  discuss this
further in
the conclusions.

Our paper is organized as follows.
In section 2 we review some renormalization group aspects
in the electric SUSY QCD
and compute the $z$ renormalization factor of the Konishi
superfield by considering its matrix elements in an external gauge
superfield.
In section 3 we apply this result to the  computation of the two-point
correlator of
the Konishi superfield at large distances and determine its conformal
dimension
at the critical point in terms of the slope of the beta function.
In section 4 we consider the magnetic SUSY QCD and construct the
duality map
for the Konishi superfield from the electric theory to the magnetic one.
We observe here a universality in the critical behaviour of the Konishi
multiplet and use duality to compute the slope of the beta function
in the strong coupling regime.
In section 5 we analyse the quantum conformal algebra at the OPE level
and point out the basic properties of the axiomatic-algebraic approach to
CFT$_4$.
We conclude with a discussion of our results and possible  
implications such as
the extension of the notion of non-Abelian duality.

\newsec{Matrix elements of currents and renormalization}

In this section we discuss the renormalization of the Konishi  
current. We first review the abelian case and then generalize the
discussion to the nonabelian case.

We start with a few preliminary comments.
In order to determine the renormalization factor of an anomalous
current it is convenient
to consider its matrix element in an external gauge field.
The renormalization
factor can be determined from that of  the corresponding anomaly.
The difference between the operator anomaly equations and the
matrix element form
of the anomaly equations was first recognized as an essential feature of
SUSY theories in ref.
\ref\puz{M. Shifman and A.Vainshtein, Nucl. Phys. {\bf B 277}  
(1986) 456.}.
A {\it natural} definition of  the Konishi current,
consistent with supersymmetry, has
been shown to lead to a one-loop form of the operator anomaly equation.
On the other hand, in ref. \ref\AJ{A.A. Anselm and A.A. Johansen,
Zh. Eksp. Teor. Fiz. Letters {\bf 49} (1989)185; Zh. Eksp. Teor.  
Fiz. {\bf 96}
(1989) 1181.}
it has been demonstrated that
the matrix elements of the divergence of anomalous axial currents  
are always
multiloop and do not depend on the  definition of the operators.
In particular in SUSY theories the  matrix elements of the
divergence of an anomalous axial
current are always proportional to the exact beta function of the
coupling constant taken at the scale of typical momenta of external  
fields.

Actually the one-loop character of the anomaly in operator form is
conventional; it is due to a particular choice of regularization.
For example, in the nonsupersymmetric case the
one-loop form of the operator equation corresponds to a ``natural''
regularization with
the help of Pauli-Villars fermions \ref\AB{S.L. Adler and W.A.
Bardeen, Phys. Rev. {\bf 182} 1517.}.
However it can be shown that for a different gauge invariant
regularization of the current this
equation can be of multi-loop form \ref\marc{M.T. Grisaru, B.
Milewski and D. Zanon,
Phys. Lett. {\bf 157 B} (1985) 174; Nucl. Phys. {\bf B 266} (1986) 589.},
\ref\AJtwo{A.A. Anselm and A.A. Johansen, Yad. Fiz. {\bf 52} (1990)
882.}.
In the simplest case of  non-supersymmetric QED such a change of the
definition of the
operators is tantamount to adding finite counterterms to the
current operator,
which are
proportional to the  current itself.

Regularization of a non-minimal type, leading to the multi-loop  operator
form of the anomaly, may seem unnatural.
However, in a supersymmetric theory this is the only possibility for
including the
anomalous axial $R$ current and the conserved energy-momentum tensor into
the same  supercurrent supermultiplet.
As for the Konishi anomaly the ambiguity of the operator equation
(its one-loop or multi-loop
character) has been demonstrated in ref. \AJ .

It is worth emphasizing that despite the operator equation ambiguity
the conditions of cancellation of  the axial anomaly are of
one-loop nature.
The point is that multi-loop corrections to the physical correlators with
insertions of
the anomalous divergence of the axial current are due to $S$-matrix-like
rescattering, while the proper anomaly is due to the one-loop
triangle
diagram.

\subsec{Abelian SUSY theories}

We first  review some of the work of  ref. \puz ,\AJ .
We  consider  supersymmetric QED
described by a real scalar (vector multiplet) superfield $V$
(the superfield strength $W_{\alpha}$  contains $F_{\mu \nu}$ and
the photino field) and two chiral matter superfields $T$ and $U$.
In this model there are two anomalous axial currents, which are the
components of two superfields, the Konishi current
$J=\bar{U} e^{-V} U + \bar{T} e^V T$ and the supercurrent
\eqn\scurrent{J_{\alpha\dot{\alpha}}={1\over
g^2}W_{\alpha}\bar{W}_{\dot{\alpha}} +
{1\over 3}D_{\alpha} (e^V T)e^{-V} \bar{D}_{\dot{\alpha}}(e^{-V}
\bar{T})+}
$$-{1\over 3}Te^{-V} D_{\alpha} (e^{V} \bar{D}_{\dot{\alpha}}(e^{-V}
\bar{T})) +
{1\over 3}\bar{D}_{\dot{\alpha}}(e^{-V}D_{\alpha}e^V{T})e^{-V}
\bar{T}  +(T\to
U ,\; V\to
-V).$$
We have absorbed the coupling constant in the definition of
the superfield
$V$.
The composite superfield $J_{\alpha\dot{\alpha}}$ contains in
addition to the
$R$-current
$R_{\mu}$, the energy-momentum tensor $T_{\mu\nu}$ and the
supersymmetry current
$S_{\mu\alpha}$.

At the classical level one has $\bar{D}^2J=0$ and $\bar{D}^{\dot{\alpha}}
J_{\alpha\dot{\alpha}}=0.$
At the quantum level the operators  $\bar{D}^2J$ and
$\bar{D}^{\dot{\alpha}} J_{\alpha\dot{\alpha}}$ are proportional to the
operator
 $W^2=W^{\alpha}W_{\alpha}$ and $D_{\alpha}W^2$ respectively.
For the supermultiplet $J$ this corresponds to the Konishi anomaly
\ref\kon{K.
Konishi,
Phys. Lett. {\bf B 135} (1984) 439.};
the anomaly in $\bar{D}^{\dot{\alpha}}J_{\alpha\dot{\alpha}}$ contains in
addition to
the
 axial $R$-current  anomaly the conformal and superconformal
anomalies.
The quantity $F_{\mu\nu}\tilde{F}^{\mu\nu}$ is contained in the
imaginary part
of the
$F$-term of the chiral superfield $W^2$.

In order to determine the anomalous dimension of the operator $J$ it is
sufficient to determine that of the operator
 $W^2$  and relate the two through the anomaly equation. For the latter
operator,
 it is sufficient to  evaluate its matrix elements in the
presence of
an external (background)
gauge field  $V_{ext} .$
 These matrix elements include multi-loop corrections due to the
rescattering of photons.

We consider the generating functional $Z$ in the presence of the external
gauge superfield
\eqn\susyZ{Z=\left(\int DV DU DT \exp [iS(V,V_{ext};U,T)]\right)_{1PI},}
$$S=\int d^4 x  \left({1\over 4g_0^2}\left[ \int d^2 \theta
W^2+h.c.\right]+
\int d^4 \theta \left[\bar{U}e^{-V}U+\bar{T}e^V T\right] \right).$$
We have omitted the gauge-fixing terms because in an abelian theory
they are
irrelevant for  what follows. Here $g_0$ is the bare coupling
constant defined
at
some ultraviolet cutoff  $\Lambda$.

The logarithmic derivative of $Z$ with respect to $g_0^2$ gives the exact
expression for the
vacuum expectation value $<W^2>$ in the external field $V_{ext} ,$
integrated over $d^4 x d^2\theta$
\eqn\vevW{\int d^4x d^2\theta <W^2> +h.c. = -4i{\partial\over
\partial(1/g_0^2)
}\log Z.}
On the other hand, $\log Z$ is the effective action
\eqn\effsusy{ Z=\exp (iS_{eff}),\;\; S_{eff}=\int d^4 x d^2\theta{1\over
4g_{eff}^2(k)}
W^2_{ext} +h.c.}
Here $k$ is a typical momentum of the external field;
$g_{eff}^2(k)$ is  the effective gauge coupling.
In eq. \effsusy \ we have kept only the terms quadratic in $W_{ext}$.
This equation is valid for $|k|^3>>|W^2_{ext}(k)|,$ in  an expansion in
powers of $|W^2_{ext}(k)|/|k|^3$.
If we now assume that the integration over $x$ and $\theta$ in eq.
\effsusy \
may be omitted, then upon comparison of eqs. \vevW \ and \effsusy \
we obtain
separate equations for matrix elements of  the chiral superfields
$W^2$ and
$\bar{W}^2$
\eqn\vevWtwo{<W^2>={\partial\over \partial(1/g_0^2) }{1\over g_{eff}^2
(k)}W^2_{ext}=
{g_0^4\over g_{eff}^4(k)}\cdot {\beta (g_{eff}^2(k))\over \beta (g_0^2)}
W^2_{ext}=}
$$={\beta (g_{eff}^2)\over \beta_1 (g_{eff}^2)}\cdot {\beta_1
(g_0^2)\over
\beta
(g_0^2)}W^2_{ext}.$$
Here  $\beta_1 =g^4 /2\pi^2$ is the one-loop
Gell-Mann-Low
function.
This equation can be easily checked by a direct computation \AJ \ to
two-loop order.

In the last equality in \vevWtwo \ the factor
$ {\beta_1  (g_0^2)/ \beta(g_0^2)}$, a function of the ultraviolet cutoff
$\Lambda$, can be removed by multiplicative
renormalization of the  composite operator $W^2$:
$$ W^2 =   {\beta_1 (g_0^2)\over
\beta (g_0^2)} W^2_{ren} $$

Thus,
$$ <W^2_{ren}> =  {\beta (g_{eff}^2)\over \beta_1 (g_{eff}^2)}  
W^2_{ext}$$
The  factor   ${\beta
(g_{eff}^2) /
 \beta_1 (g_{eff}^2)}$ depends on the external momentum and therefore
determines the anomalous dimension of the operator $W^2_{ren}$, and  
hence,
by the anomaly equation,   as we discuss below,
of $J$.

As  described in ref. \marc , in the dimensional
regularization scheme
there is a natural definition of the operator
$J_{\alpha\dot{\alpha}}$ which
contains a conserved
component of spin 2  corresponding to the energy-momentum tensor.
By an explicit computation it has been shown \marc \ that the operator
anomaly equation for $\bar{D}^{\dot{\alpha}}J_{\alpha\dot{\alpha}}$ has a
multi-loop
character and is
proportional to the exact multi-loop beta function $\beta$.
Since the operator $J_{\alpha\dot{\alpha}}$ contains conserved
currents, it
is finite and needs no renormalization.

For the definition of the   {\it renormalized} Konishi superfield   
operator $J$
we
have more
freedom,
since it does not contain any conserved current.
For example we can accept a definition \puz \ of $J$ by the
requirement
that
it coincides with the bare current at the   ultraviolet cut-off  
$\Lambda$.
In this case all radiative
corrections to the matrix element of this operator in the presence
of external
fields
vanish at $k^2=\Lambda^2$, where $k^2$ is a typical external momentum (we
assume
that the differences of momenta of individual external legs are
negligible as
compared to $k^2$).
Matrix elements of such an operator will be  proportional to a factor
$z_J(g^2(k))/z_J(g_0^2)$, with  $z_J(g)$ to be determined below.

To analyse the  renormalization of   $J$ one has to
consider a mixing of
the operators $D_{\alpha}\bar{D}^2 J$,
$\bar{D}^{\dot{\alpha}}J_{\alpha\dot{\alpha}}$
and $D_{\alpha}W^2$.
Similarly to the case of non-supersymmetric QED it can be shown \AJ
\ that the
operator
$D_{\alpha}\bar{D}^2 J$ can be renormalized by a multiplicative
 factor
$z_J(g_0^2)$
which does not depend explicitly on the parameter $\Lambda ,$
$(D_{\alpha}\bar{D}^2 J)_{ren}=
z_J(g_0^2)D_{\alpha}\bar{D}^2 J$.
One also finds that the following operators
are invariant under the renormalization group
\eqn\opanSUSY{\bar{D}^{\dot{\alpha}}J_{\alpha\dot{\alpha}}+ {\beta
(g_0^2)\over
6g_0^4}D_{\alpha}W^2=c_1 ,\;\;
D_{\alpha}\bar{D}^2 J-{\beta_1 (g_0^2)\over g_0^4}D_{\alpha}W^2=c_2,\;\;
{\beta (g_0^2)\over \beta_1 (g_0^2)}D_{\alpha}\bar{D}^2 J=c_3,}
where $c_{1,2,3}$ are constants, $\beta_1$ is the one-loop beta function,
and the other operators and $g$ are taken at a renormalization point $m$.
The constants $c_{1,2,3}$ can be fixed by considering the limit
$g\to 0$, when obviously
$c_1=c_2=0$ as a consequence of the one-loop approximation.
Thus the corresponding operator equations read \puz \
\eqn\opeqSUSY{\bar{D}^{\dot{\alpha}}J_{\alpha\dot{\alpha}}=- {\beta
(g_0^2)\over
6g_0^4}D_{\alpha}W^2,\;\;\;
D_{\alpha}\bar{D}^2 J={\beta_1 (g_0^2)\over g_0^4}D_{\alpha}W^2 .}

Finally, for  for renormalized matrix elements of the operators
 in the external field $V_{ext}$ one obtains \AJ \
\eqn\msusy{<\bar{D}^{\dot{\alpha}}J_{\alpha\dot{\alpha}}>=-{\beta
(g^2)\over 6g^4} D_{\alpha}W^2_{ext},\;\;
<\bar{D}^2 J >_{ren}={\beta (g^2)\over g^4}W^2_{ext},}
where $g=g(k)$ is the effective coupling constant.

Comparing  \opeqSUSY  , \vevWtwo  \ and  \msusy  \ one verifies
that indeed
$J_{\alpha \dot{\alpha}}$ is not renormalized, whereas  the
renormalization
factor for the
Konishi current is
$z_J(g_0^2)=\beta(g_0^2)/\beta_1(g_0^2).$:
\eqn\jren{ J_{ren} = {\beta(g_0^2) \over \beta_1(g_0^2)} J.}
Equivalently, as established in ref. \puz , $(1-\gamma) J$ is  
renormalization
group invariant,
$\gamma$ being the anomalous dimension of the chiral superfields  
$T$ and $U$.

\subsec{Non-abelian SUSY theories}

We  now  generalize  to the non-Abelian SUSY theory.
For definiteness we shall consider  SUSY QCD with $SU(N_c)$
gauge group and
 $N_f$ flavours of
 chiral superfields $Q_i$ and $\tilde{Q}^i$, $i=1,...,N_f$, in the
fundamental
representation.
One can define  the supercurrent superfield $J_{\alpha
\dot{\alpha}}$ and
the Konishi
superfield $J=\sum_i ( \bar{Q}_i e^V Q_i + \tilde{Q}_i e^{-V}
\bar{\tilde{Q}}_i) $
 which are singlets under the flavour global
$SU(N_f)\times SU(N_f)$
group.
These currents are anomalous.
One can follow the above analysis to determine the matrix elements
of these
currents.
The analysis here is however less trivial because the external field is
charged with respect to the gauge group.
Therefore the matrix elements depend on the gauge fixing \AJ \
and hence they are not simply of the form given
in eqs. \msusy .
However in an appropriate kinematical regime \ref\nrjoh{A. Johansen,
Nucl. Phys. {\bf B 376} (1992) 432.}
 \ref\MAnr{M.A. Shifman and A.I. Vainshtein, Nucl. Phys. {\bf B 365}
(1990) 312.}
which corresponds to a limit of
vanishing momentum of the current these matrix elements reduce to the
form of
eqs. \msusy .
In what follows we shall consider this particular kinematics since,
as we shall
see,
it is relevant for computation of the two-point correlators
discussed below.

By differentiating the 1PI effective action $S_{eff}$
with respect to the bare coupling constant
one gets similarly to the case of the abelian theory the following
equation
\eqn\WNAsusy{<{\rm Tr} \; W^2>= {\beta (\alpha)\over \beta_1
(\alpha)}\cdot
{\beta_1
(\alpha_0)\over \beta (\alpha_0)}\cdot
{1\over 1-\alpha_0 N_c/2\pi} {\rm Tr} \; W^2_{ext} .}
We have adopted a definition of the operator $W^2$ at a  scale $m$ (below
$\Lambda$) which agrees
with that of
ref.
\puz \
and is responsible for the factor $1-\alpha_0 N_c/2\pi$ (see below).
The total NSVZ beta function \ref\NSVZ{V. Novikov, M. Shifman, A.
Vainshtein
and V. Zakharov, Nucl. Phys. {\bf B 229} (1983) 407;  {\it ibid}
{\bf B 229}
(1983) 381;
Phys. Lett. {\bf B166} (1986) 329.} \ reads
\eqn\bfunction{\beta (\alpha)=-{ \alpha^2\over 2\pi} \cdot
{3N_c-N_f +N_f\gamma
\over 1-\alpha N_c/2\pi},}
where $\alpha=\alpha(k)=
g(k)^2/4\pi$ is the effective coupling at the scale of external
momenta.
Here $\gamma$ is the anomalous dimension of the matter
superfields
\eqn\andim{\gamma (\alpha)=-{d\log Z\over d\log m}=
-{N_c^2-1\over 2N_c}\cdot {\alpha \over \pi} + O(\alpha^2).}
The factor $1/ (1-\alpha_0 N_c/2\pi)$ appears because of the
one-loop form
of the Wilsonian action  \puz
 \eqn\effact{S=\int d^4x  \left( {1\over 2g_0^2} +b_1\log {\Lambda\over
m}\right)
{\rm Tr} [W^2]_F +
{Z\over 4} \int d^4x   [\bar{Q}_ie^{V}Q_i
+\tilde{Q}_ie^{-V}\bar{\tilde{Q}}_i]_D }
where $b_1$ is the first coefficient of the expansion of the beta
function in powers of the
coupling constant.
The derivative of the effective action \effact \ with respect to
$1/[g_0^2]$  (in the notation of ref. \puz ) and
hence to $\log \Lambda$
gives the additional factor in  \WNAsusy. The operator ${\rm Tr}
\;W^2$ used
in eq.
\WNAsusy \ differs by the constant factor $1/(1-\alpha_0 N_c/2\pi)$
from  the bare operator ${\rm Tr}\; W^2$
at the renormalization scale $m=\Lambda$, according to the
definitions used
in ref. \puz .

The analysis of the anomaly operator equations is also more subtle
\puz \ and
gives
 (at a renormalization
scale $m$)
\eqn\NAsusy{\bar{D}^{\dot{\alpha}}J_{\alpha\dot{\alpha}}=
-{3N_c-N_f +N_f\gamma
(\alpha_0)\over 48\pi^2}D_{\alpha}W^2 ,\;\;
D_{\alpha}\bar{D}^2 J={N_f\over 2\pi^2}D_{\alpha}W^2.}
Thus, the anomaly of  the $J_{\alpha\dot{\alpha}}$ current is
proportional to the
numerator of the NSVZ beta function \bfunction .
Consequently,  from eqs. \WNAsusy \ and \NAsusy , the
renormalized matrix
elements of the currents read
\eqn\msusyNA{<\bar{D}^{\dot{\alpha}}J_{\alpha\dot{\alpha}}>={\beta
(\alpha)\over
24\pi
\alpha^2}  D_{\alpha}W^2_{ext},\;\;
<\bar{D}^2 J_{ren} >={N_f\over 2\pi^2}\cdot
{\beta (\alpha)\over \beta_1 (\alpha)}W^2_{ext},}
which is a generalization of eqs. \msusy  \  to the nonabelian case.
This shows that the $z$ factor for the Konishi superfield
is given by $z( \alpha_0 )=\beta (\alpha_0)/\beta_1 (\alpha_0)$.
In the above  equation
$$ J_{ren} =  (1-{\alpha_0 N_c \over 2\pi}) z(\alpha_0)J .$$
This equation deserves a short comment.
It is well-known that in an asymptoticaly free theory
the renormalization factor 
$Z_J(\log \Lambda/\mu,\alpha_0)
={\rm exp}\left(-\int_{\alpha_0}^{
\alpha(\log \Lambda/\mu,\alpha_0)}\gamma_J(\alpha')d\alpha'
/\beta(\alpha')
\right)$ of an anomalous current $J$
admits a smooth limit $Z_J(\alpha_0)={\rm exp}\left(\int^{\alpha_0}_0
\gamma_J(\alpha')d\alpha'
/\beta(\alpha')
\right)$ when the cut-off is removed \ref\Collins{See for example
J.C. Collins, Renormalization, Cambridge University Press,  
1985, par. 13.6}, since
$\gamma_J$ vanishes at the one-loop order. 
The finite factor that we have determined
can be seen as this limit.

 \newsec{Correlators of anomalous currents at large distances}

In this section we examine the two-point function of  the Konishi  
current at large
distances, and determine its anomalous dimension.

The scaling dimension of an operator at the conformal fixed point
is  the sum of its classical ($d_0$) and
anomalous ($\gamma$) dimensions at the critical values of the
coupling constants.
To determine the scaling dimension of a local operator $O$ it is
sufficient to
consider
a two-point  correlator $<O(x) O(y)>$ at large distances $|x-y|\to
\infty .$
The Callan-Symanzik equation
$$\left(|x-y|{\partial\over \partial |x-y|} +2d_0 +2\gamma (\alpha )+
\beta (\alpha ){\partial\over \partial \alpha }\right) <O(x) O(y)>=0 ,$$
implies that the correlator has the following form:
$$<O(x) O(y)>=(z(|x-y|))^2 \phi (\alpha (|x-y|))$$
Here $z$ is a renormalization factor for the operator $O$,
$\alpha (|x-y|)$ is the
running coupling constant at the scale $1/|x-y|$ and  $\phi (\alpha )$ is
an unknown
function.
If the function $\phi (\alpha )$ has a non-vanishing smooth limit at the
critical
value $\alpha_*$
of the coupling constant then the leading behaviour of such a
correlator is
$$<O(x) O(y)>\sim {1\over |x-y|^{2d_0+2\gamma_*}},$$
where $\gamma_*$  is  the anomalous dimension at the critical
point.
Having established that the renormalization factor for  the Konishi  
current is
expressible in terms of the beta function, it will follow
 that its scaling dimension at the critical point is
determined  by the slope of the beta function.

To fix the ideas, we review  the  computation of the
scaling dimensions of
fundamental fields in the electric formulation of
 SUSY QCD.
More specifically we consider the $SU(N_c)$ supersymmetric gauge theory
with $N_f$ flavours  $Q_i$ and $\tilde{Q}^i$, $i=1,...,N_f$, of
chiral quark superfields in the
fundamental and anti-fundamental representation of the gauge group
respectively.

We consider the propagator of  the {\it scalar} components of quark  
superfields
at large distances.
We will  verify  that  their conformal dimension at the conformal
fixed point
coincides with $3R_Q/2$, where
$R_Q=1-N_c/N_f$
is the $R$ charge \emd .
By standard arguments  for elementary fields
 $\phi (\alpha (|x-y|) \rightarrow |x-y|^{-2}$ and
 we have, for large
$|x-y|$,
\eqn\prop{<Q_i (x) \bar{Q}^j (y)> =\delta^j_i  z_Q (\alpha(x-y))
\cdot {1\over
|x-y|^2}.}
Here $[z_Q(\alpha)]^{1/2}$ is the wave renormalization factor for
the field
$Q_i$.
At large distances the running gauge coupling $\alpha(x-y)$ flows
to the fixed
value
$\alpha_*$ which corresponds to the conformal fixed point.
In this regime we have \ksv \
\eqn\Zq{z_Q(\alpha)=z_Q(\alpha_0) \exp \left(-\int^{\alpha}_{\alpha_0}
d \alpha {\gamma_Q (\alpha )\over \beta(\alpha )}\right)\sim
(\alpha_*-\alpha)^{-\gamma_Q(\alpha_*)/\beta'(\alpha_*)}}
where $\gamma_Q(\alpha)$ is the anomalous dimension of the
quark field
$$\gamma_Q(\alpha)=-{d\log z_Q(\alpha(m))\over d\log m }.$$

Near the critical point the beta function is supposed to have a simple
zero, i.e. $\beta(\alpha)=\beta'(\alpha_*)(\alpha-\alpha_*).$
This means that the correlators are power-behaved at the critical
point. Higher
zeroes would
instead produce logarithmic behaviour. Solving the equation
$d\alpha(m)/d\log
m=\beta (\alpha(m))$ one gets
\eqn\gcr{\alpha_*-\alpha=|x-y|^{-\beta'(\alpha_*)}\;\;{\rm at} \;\;
|x-y|\to
\infty .}
Here $\alpha$ is taken at the scale $1/|x-y|$.
Substituting  this expression into eq. \prop \ we get
\eqn\propAs{<Q_i (x) \bar{Q}^j (y)> =\delta^j_i  {1\over |x-y|^{2+
\gamma_Q (\alpha_*)}}.}
By using now the fact that $\gamma_Q(\alpha_*)=-(3N_c-N_f)/N_f$
we see that the conformal dimension of the quark field at the fixed
point is
$d_Q= 1+ {1\over2}\gamma_Q (\alpha_*)=1-(3N_c-N_f)/2N_f=3(1-N_c/N_f)/2.$
The latter value coincides with the value
$d_Q=3R_Q/2$ which is required by the superconformal algebra.

Turning to the  operator $J$ whose  $z$ factor was determined in  
the previous
section, we consider, at the component level, the two-point  
correlator of its
axial vector component
 $a_{\mu} \sim [\bar{D}_{\dot{\alpha}} , D_{\alpha} ] J|_{\theta=0}$ .
The Callan-Symanzik equation  leads to the following form for
the  correlator :
\eqn\KK{<a_{\mu}(x) a_{\nu}(y)>=\left({\beta (\alpha)\over
\beta_1(\alpha)}\right)^2 \cdot \left(
{\phi_1 (\alpha)g_{\mu\nu}\over |x-y|^6}+
{(x-y)_{\mu}(x-y)_{\nu}\phi_2 (\alpha )\over |x-y|^8}\right) .}
The presence of the factor $(\beta
(\alpha)/\beta_1(\alpha))^2$
is our main observation.
The functions $\phi_{1,2} (\alpha)$ are not determined by the
Callan-Symanzik equation.

We consider the large distance limit of this correlator.
At large distances the running gauge coupling $\alpha(|x-y|)$ flows
to the fixed
value
$\alpha_*$ which corresponds to the conformal fixed point.
The factor $(\beta (\alpha)/\beta_1(\alpha))^2\to 0$ at $|x-y|\to \infty$
because $\alpha$ goes to the critical value $\alpha_*$.
In general the functions $\phi_{1,2} (\alpha)$ could have zeros or
poles at the
critical
point  but, as we shall discuss below, this is not  
generically
the case.
Using   $\beta(\alpha)=\beta'(\alpha_*)(\alpha-\alpha_*)$ and
substituting the  expression from  eq. \gcr  into eq. \KK \ we get
\eqn\Kcr{<a_{\mu}(x) a_{\nu}(y)>\sim {1\over
|x-y|^{6+2\beta'(\alpha_*)}} .}
Thus, the anomalous dimension of the Konishi current is given by  
the slope of
the beta-function.

Our result  can be easily checked in perturbation theory. We outline the
calculation in the Appendix.

We discuss now the behaviour of the functions  $\phi_{1,2}$.
In  perturbation theory we have \AJ  \
$\phi_{1,2}  (\alpha)=1+O(\alpha^2).$
Therefore at least for the models with $\alpha_*<<1$ the functions
$\phi_{1,2} (\alpha)$ have a smooth non-vanishing limit at $\alpha\to
\alpha_*$~\foot{Note that the coefficients in the expansion of
the functions
$\phi_{1,2} (\alpha)$ are not singular
at $(3N_c-N_f)/N_c\to 0$ because these terms are finite without any
additional
renormalization of the correlator.}.
Thus in these theories the scaling behaviour of the correlator
is determined only by the factor $(\beta
(\alpha)/\beta_1(\alpha))^2\to 0$.
(Note however that
the value of $\phi_{1,2}  (\alpha)$ together with the above factor
determine
the values of some of central charges of the superconformal algebra
in section 5.)

For the fixed point in the strong coupling regime we do not have any
argument for a smooth non-vanishing limit of
$\phi_{1,2}  (\alpha)$ at $\alpha \to \alpha_*$.
One could imagine  that a factor $\sim \exp (-M|x-y|)$ appears due
to nonperturbative effects, where $M$ stands for a mass parameter
corresponding
to the most relevant intermediate state.
If this were so  it would mean that there are only massive
intermediate states $|n>$ in the  correlator
$<a_{\mu}(x) a_{\nu}(y)>$.
However in the non-Abelian Coulomb phase the quark fields are massless
as  manifested in the power-like behaviour of
the two point correlator $<Q_i (x) \bar{Q}^j (y)>$.
This suggests that the spectrum of  intermediate states in the
$<a_{\mu}(x) a_{\nu}(y)>$ channel has no mass gap, and hence the
power-like
behaviour of this correlator may remain in the infrared even in the
strongly coupled regime.
Note that this power-like behaviour could be also modified by softer
factors $\phi_{1,2}$ which have a zero or a pole at $\alpha=\alpha_*$,
i.e.  $\phi_{1,2}\sim
(\alpha_*-\alpha)^{a}$ where $a$ is a constant.
This would change the conclusions which we  have presented.
However by continuity one can argue that this is not the case.
Indeed we assume that there is no singularity in the correlator
with respect to
$N_c$ and $N_f$ in the conformal window.
Then the exponent $a$  depends continuously on the numbers
$N_c$ and $N_f$.
Therefore such a non-trivial behaviour of the factors $\phi_{1,2}$
would extend also to the domain of the
weak coupling regime in the space of parameters  $N_c$ and $N_f$.
However we saw that the factors $\phi_{1,2}$ do not have any zeros
or poles in
this domain.
Therefore we may expect at most that a zero or a pole could appear
only outside
the
conformal window $0<{3N_c-N_f\over N_f}<1$.
Below  we shall argue that this is indeed the case  for
the theory in the confining phase, i.e. at $3N_c/2\geq N_f$.

It is instructive to comparer  the above result
with the infrared behavior of  non-asymptotically free theories with
$N_f>3N_c$.
Assuming that such a theory flows into a free theory, which
corresponds to
$\alpha \to 0$, we see that the factor $\beta/\beta_1\to 1$ in the
infrared.
The correlator of Konishi current at large distances has an integral
power-like behaviour
which implies
that all currents have a canonical dimension as expected in a free
theory.

So far we discussed perturbative aspects of the infrared behavior  
of SQCD.
The nonperturbative effects are due to instantons and,
hence, are proportional to $\exp (-2\pi/\alpha)$.
In a non-asymptotically free theory, the coupling constant in
the infrared is small, and hence instanton effects are absent.
In a theory which flows to the non-trivial conformal fixed point
in the infrared these effects are $\sim \exp (-2\pi/\alpha_*)$.
Therefore at least in the regime when the critical value $\alpha_*$ of
the coupling constant
is small, $\alpha_*<<1$, the instanton effects are exponentially
suppressed and, hence, are not important.

We believe that these effects are not important for computation of
the conformal dimensions even for strongly coupled conformal points.
In particular 
we saw that there are no instanton corrections to the quark
conformal dimension.
This follows from the fact that  the power-like behaviour of the quark
propagator at
large distances is determined by the perturbative beta function.

In a confining theory, e.g. N=1 SYM,  the perturbative beta function has
a pole (in the theories with a non-trivial fixed point such a pole
is screened by a zero of the beta function).
For N=1 SYM the large distance
behaviour of the correlators corresponds to approaching  the pole
in the beta function, $\alpha\to 2\pi/N_c$.
Therefore naively the correlator of two Konishi currents at points with
coordinates $x$ and $y$ is proportional to
$${1\over |x-y|^4\log (m |x-y|)}$$
in the limit $|x-y|\to 1/m,$ where $m$ is a scaling parameter.
Such a behaviour is physically meaningless.
Therefore the instanton effects and, hence, the factors $\phi_{1,2}$ are
crucially important to reproduce a correct behaviour of the correlator.
To screen a pole at $|x-y|\to 1/m$ the functions $\phi_{1,2}$ must
have a zero at
$\alpha\to \alpha_*=2\pi/N_c$ as mentioned above.

\newsec{Konishi current in the magnetic theory}

We  consider  now the magnetic formulation
of the theory \emd , which has the $SU(N_f-N_c)$ gauge group, $N_f$
flavours of  dual quarks $(q^i,\tilde{q}_j)$
in the fundamental and anti-fundamental representations, respectively,
and
the meson field $M^i_j$ in the $(N_f,\bar{N_f})$ representation of the
flavour $SU(N_f)\times SU(N_f)$ group.
In constrast to the electric formulation this theory has a superpotential
$S=M^i_j q^i \tilde{q}_j$.

We discuss the determination of  the magnetic counterpart of the Konishi
current of the electric theory.
The conservation of the Konishi current in the electric theory is  
violated by
quantum effects.
{\it A priori} there is no reason for the corresponding  current in the
magnetic
formulation of the theory
to be conserved even at the classical level.
Its conservation may be violated due to the superpotential.
As we shall see below the conservation of the magnetic
Konishi current
is spoiled both by the superpotential and the gauge anomaly.

To analyse the situation we consider an  extension of the electric
theory which
includes an additional
chiral superfield $X$ in the adjoint  representation of the gauge group.
Such an extension was analysed in ref. \kutasov .
In this model the chiral superfield $X$ of the electric theory has a
superpotential
$$S_{el}=s{\rm Tr} X^{k+1},$$
where $k$ is a positive integer and $s$ is a coupling constant.
We shall focus on $k=2$, which gives the only interesting
renormalizable theory that flows to a non-trivial superconformal theory
(for $N_c/2< N_f < 2N_c$, see ref. \kutasov).
On the magnetic side the theory includes the meson fields
corresponding to
\eqn\meson{(M_h)^i_{\tilde{i}}=\tilde{Q}_{\tilde{i}}X^{h-1} Q^i;\;\;\;
h=1,...,k .}
It also includes an additional
chiral superfield
$Y$ in adjoint representation of the dual gauge group $SU (kN_f-N_c)$.
The magnetic superpotential reads
$$S_{mag}= \tilde{s} {\rm Tr} Y^{k+1} + \sum_{h=1}^k t_h
M_h\tilde{q}Y^{k-h}q,$$
where  $\tilde {s}$ and $t_h$ are coupling constants.
The precise duality map (for primary chiral operators)
between these two formulations of the theory
(and its deformations) has been constructed in ref. \kutasov .
Here we are interested in a particular aspect of such a map.
The strategy that we shall employ is to construct  nonanomalous Konishi
currents
for the  dual versions of {\it this} theory, including the fields
$X$ and $Y$,
and
subsequently, by introducing  large mass terms for these
fields, recover the original  theory and corresponding currents in
a low-energy
limit. It is convenient to study the correspondence of the electric and magnetic
theories
at the level of the Wilsonian action.

We consider the Kutasov theory \kutasov  \ at its critical point
(so that in
particular the coupling constants have their critical values for
{\it this } theory $\alpha = \alpha_\sharp$, etc.) and
 perform a non-anomalous $U(1)$ phase transformation of the
Kutasov electric theory fields
$Q$, $\tilde{Q}$ and $X$ with charges $q_Q=q_{\tilde{Q}}$ and $q_X$
respectively.
The absence of an anomaly requires that $N_f q_Q +N_cq_X=0$
($1/2$ for each fundamental representation, $N_c$ for each adjoint;
only the
matter sector is relevant, since the Konishi current does not contain
the gauge sector).
In the magnetic case, the corresponding condition  gives
$N_fq_q=(N_c-N_f)q_Y$.
The lagrangian of the electric theory is invariant under such a
transformation provided
we transform the coupling constant $s$ with the charge $q_s=-(k+1) q_X$.

We note  that the magnetic
lagrangian should be also invariant under the transformation
corresponding to that of the electric fields provided the
coupling constants are also transformed in an appropriate way.
By postulating that the transformation properties of the meson fields are
prescribed by the identification \meson \
and due to the holomorphy of the Wilsonian action, we can easily get
that the charges of the couplings  $t_h$ and $\tilde{s}$ are given by
\eqn\tcharges{q_{t_h} =(2N_c/N_f +1-h)q_X + (h-k+2(N_f-N_c)/N_f )q_Y ,}
$$q_{\tilde{s}}= -(k+1)q_Y .$$
By considering a matching of deformations of these theories
\kutasov \ it is easy to see that
the charges of the couplings $t_h$ do not change under a
deformation $N_f\to N_f-1$.
This condition fixes uniquely  $q_X=q_Y$, and hence
$$q_{t_h}=-q_s{3-k\over k+1} , \;\;\; q_{\tilde{s}}=q_s .$$

As a byproduct of the above considerations, writing
$t_j=t_j(s)=s^{q_{t_j}/q_s}=s^{k-3\over k+1}$ to denote the
dependence of the
magnetic coupling on the electric one\foot{Our normalization is different
from that of  \kutasov ; our $t_j$ corresponds to their $s/\mu^2$.}
we see that holomorphy in the electric
parameter $s$ means that the coupling $t_j$ is singular
in the limit $s\to 0$ at $k<3$, for example in the theory with
$S_{el}= s{\rm Tr} X^3$.
This suggests that new additional massless bound
states appear in the magnetic theory in this limit. This means that
for duality
to work
in the case $s=0$ we have to add a new (massless) field.

 In the electric theory we define a non-anomalous current which is a
linear combination
of the Konishi current  and a current constructed from  the field $X$
(the conservation of this current is broken by the superpotential
$S_{el}$)
\eqn\KKU{K_{el}= \sum_i\left[Q^+_ie^VQ_i+\tilde{Q}_i
e^{-V}\tilde{Q}_i^+\right]
-
{N_f\over N_c} {\rm Tr} X^+e^V Xe^{-V}.}
The corresponding current in the magnetic theory is defined by the $U(1)$
transformation with the charges found above.

In order to determine the duality map for the Konishi operator in
the original SQCD
we  add now a mass term  $m {\rm Tr} \;X^2$ and, via integration
over $X$ (or, equivalently, by the Appelquist-Carazzone  theorem)
go back to the  minimal  theory with fields $Q,$ $\tilde{Q}$ on the  
electric
side.
Also, by the duality map for chiral operators constructed in ref.
\kutasov \  {\it at the critical point of the Kutasov theory},     
$Y$  will
have a mass term $m {\rm Tr} \;Y^2$  and we  will then reproduce
the original theory in the low-energy limit with the fields $q$,
$\tilde{q}$ and $M$ on the magnetic side.
Moreover, to reproduce precisely the initial theory,
we have to choose a phase in which the gauge group is broken down to
$SU(N_f-N_c)$.
Note that the integration over the heavy field $X$ (and $Y$ on the
magnetic side) leads to a non-critical minimal SQCD because,
in particular, the coupling constants do not have  their critical values,
e.g. $\alpha = \alpha_\sharp \neq \alpha_*$.
Alternatively, we can think of the heavy fields (with masses of  
order $m$)
as regulators of the minimal SQCD away from its fixed point.
The parameter $m$ is the only scale in the Kutasov theory and
plays the r\^ole of UV cutoff in this resulting non-critical theory.
The effective Konishi operators in this effective non-critical SQCD
will be formally obtained by dropping  out the heavy fields  
appearing in the
non-anomalous Konishi operators of the Kutasov model.
The resulting Konishi operators should be thought of as
bare operators defined at this scale.
Turning  things around, the Konishi operators in the deformed Kutasov theory
may be thought as particularily regularized operators of the  
minimal SQCD.

More specifically, after dropping the fields $X$, $Y$, on the  
electric side
the Konishi current
$J_{el}$ has its usual form, and is expected to flow in the  
infrared to the
corresponding current in the critical minimal theory.
Its magnetic counterpart
is obtained  by taking into account that $q_M=2 q_Q$ for $h=1$ in  
eq. \meson .
(For $h>1$ $M_h$ disappears in the low-energy limit, since it  
contains $X$).
It can be written as
\eqn\magkon{J_{mag}=2J_M +{N_f-N_c\over N_c} J_q}
where
$$J_M={\rm Tr} \bar{M}M, \;\;\; J_q=
\sum_{f=1}^{N_f} \bar{q}_fe^V q^f +\tilde{q}_f e^{-V}\bar{\tilde{q}} ,$$
The conservation of this current is broken both by the gauge anomaly  and
the superpotential $S=\lambda M^i_jq_i\tilde{q}^j$, where
$\lambda \sim  t_k <Y^2> $ is a coupling constant.
As mentioned above, the operator \magkon \ is defined in the theory  
formulated
at the scale $m$.
At this scale the coupling constants in the magnetic SQCD are equal
to their values at the critical point in the theory with field $Y$  
present,
$\alpha=\alpha_{\sharp}$, $\lambda=\lambda_{\sharp}$.
Below the scale $m$ the low energy theory is
the usual non-critical minimal SQCD.
Thus the operator \magkon \  is defined in the magnetic
minimal  SQCD away from criticality
on a particular RG trajectory in  the  $(\alpha ,\lambda)$ phase diagram.

In order to describe the duality map for the Konishi operators
in the critical SQCD we have to analyse their  RG flow towards the  
infrared.
Equivalently, we are interested in the large distance ($>>1/m$)
behavior  of the correlators which is governed by the critical SQCD.
Note that in the above discussion we have {\it identified}
the electric and magnetic
Konishi operators in {\it the Kutasov critical theory} in the  
presence of the
above
mass deformation.
Therefore their behaviour at any distance will be identical.
This also means that the magnetic
Konishi current is multiplicatively renormalizable and does not mix  
with other
operators in the effective off-critical magnetic theory as soon as  
this is true
for the
Konishi current on the electric side.
Therefore the magnetic Konishi operator can be represented
as a linear combination of renormalization group invariant operators in
the magnetic version of an  off-critical  {\it minimal} SQCD
\eqn\comb{J_{mag}= AJ_1+BJ_2 .}
Here $J_1$, $J_2$ form a basis of RG invariant operators
which diagonalize the  matrix of anomalous  dimensions and $A$, $B$ are
constants which depend on the critical values of the coupling  
constants $\alpha
= \alpha_{\sharp}$,
 $\lambda = \lambda_{\sharp}$  in the Kutasov theory, and on the  
scale $m$.
Such a representation is convenient for a consideration
of the  infrared behaviour near the conformal fixed point.

We will now analyse the matrix of anomalous  dimensions of  component
axial currents in the magnetic SQCD.
In the off-critical effective low-energy minimal SQCD theory, one  
of the RG
invariant operators is the non-anomalous conserved
$R$ current defined in ref. \emd  \
which however does not mix with the Konishi current
since these are components of the supermultiplets with different  
superspins.
Two other currents are those which are not conserved due to the
anomaly and the superpotential.
It is convenient to consider  the current
$J_{W}=J_q-2J_M$ whose (super)divergence is
given by the anomaly and the current $J_{sp}=J_M$ with divergence
proportional to the superpotential so that
$J_{mag}={N_f-N_c\over N_c} J_W + 2{N_f\over N_c} J_{sp}$.
These two currents are mixed under RG flow.
One RG invariant combination is given by
$$J_1={2 N_f-3N_c+N_f\gamma_q\over 48\pi^2}J_{W}+
\beta_{\lambda} ~J_{sp}$$
which remains invariant under RG flow because its $D_{\alpha}\bar{D}^2$
divergence is proportional to the anomaly of the supercurrent
  $J_{\alpha\dot{\alpha}}$
\ksv ,  \ref\md{M.T. Grisaru, B. Milewski
and D. Zanon, Nucl. Phys. {\bf B 266} (1986) 599.}
\eqn\scamag{\bar{D}^{\dot{\alpha}} J_{\alpha\dot{\alpha}}=
-{2 N_f-3N_c+N_f\gamma_q\over 48\pi^2}D_{\alpha}\bar{D}^2
{\rm Tr}\; W^2_{mag} + \beta_{\lambda} D_{\alpha}\bar{D}^2 S.}
Here $\beta_{\lambda}=
\lambda (\gamma_M+2\gamma_q)/2$ is the beta function of $\lambda$, and
$\gamma_q$ and $\gamma_M$ are anomalous dimensions
of the fields $q,\tilde{q}$ and $M$ respectively.
It is easy to see that this combination is different from \magkon .

Another RG invariant combination, $J_2$,  is less trivial to obtain.
In order to determine it we have to actually  compute the matrix of
anomalous
dimensions.
This can be done  by the same procedure we used in section 2 for the
electric theory
by taking the derivatives of the effective action with respect to
the bare coupling constants.
As a result the $z$ matrix is determined uniquely by the beta
functions via
suitable differential equations.
After tedious calculations (we do not present them here)
one can find that the entries of this $z$ matrix are
proportional to linear combinations of beta functions in the  
infrared limit.
We can rewrite the expression \magkon \ in the form \comb\
with $A=A(\alpha_{\sharp},  \lambda_{\sharp} , m)$,   
$B=B(\alpha_{\sharp},
\lambda_{\sharp} , m)$ uniquely fixed by the beta functions.
The values of $A$ and $B$ have well defined limits at $m\to \infty$ which
are both non-zero.
These  particular values  will  not be important for us.

At this point we wish to make a comment.
In an off-critical theory even operators defined in
an RG invariant way depend on a particular RG trajectory.
This follows from the fact that a correlator of such operators  
depends on the
value of
$\Lambda_{QCD}$ which labels RG trajectories.
The construction above used a particular RG trajectory
(from the initial critical theory with $X$ and $Y$ integrated out)  
into the
critical SQCD. The information about the particular  trajectory is   
encoded in
the parameters  $ \alpha_{\sharp}$,  $\lambda_{\sharp}$  , and $m$.  
One can
consider a more general flow along an arbitrary RG trajectory. In  
this case the
correlators of $J_{1,2}$ and hence the operators themselves depend on the
chosen trajectory because the running coupling constants do. On the  
other hand,
the operator $J_{mag}$  at the critical point  (of the minimal  
SQCD) should not
depend on the choice of trajectory in the $(\alpha , \lambda )$  
plane. Eq.
\comb \  still holds, with the
coefficients $A$ and $B$ compensating the dependence of the ooperators
$J_{1,2}$ on the trajectory. Thus these coefficients, specifying a  
particular
trajectory, are now to be interpreted as coordinates on the space of
non-critical renormalizable deformations of the magnetic SQCD.

As a byproduct of the above discussion one can obtain the
magnetic dual description of the operator ${\rm Tr} \;W^2_{el}$.
Indeed our identification of the magnetic counterpart for the  Konishi
supermultiplet implies that the operator ${\rm Tr} \;W^2$ in the
electric theory matches on the magnetic side
with  a linear combination of the operator ${\rm Tr} \;W^2$ and the
superpotential, which is proportional to the superdivergence of the
operator $J_M$.
Thus the duality map that we get for the anomaly multiplet
seems to be different from that of  ref. \exam \ where it has been
suggested
${\rm Tr} \;W^2_{el}=- {\rm Tr}\; W^2_{mag}$.
By our analysis this relation is  modified as follows:
after the integration over the heavy fields in the Kutasov model we have
(away from criticality of the minimal SQCD theory, at the scale $m$)
$J_{el}\sim J_{mag}={N_f-N_c\over N_c} J_W + 2{N_f\over N_c} J_{sp}$,
so that
${\rm Tr} W^2_{el}\sim- {\rm Tr} W^2_{mag} +{N_f\over N_c}
({\rm Tr} W^2_{mag} +2S) .$

We now compute the anomalous dimension of the magnetic Konishi
current.
Consider the two
points correlator $<a_{\mu} (x) a_{\nu} (y)>$ at large distances, where
$a_{\mu}$ is the spin 1 component of $J_{mag}$.
This correlator is essentially determined by the matrix of anomalous
dimensions,
and hence by the $z$ factor which takes into account the mixing of the
operators
${\rm Tr} \;W^2_{mag}$ and $S$.
As mentioned above the entries of $z$ matrix renormalization factor  
are given
by linear combinations of the beta functions.
Therefore the correlator  $<a_{\mu} (x) a_{\nu} (y)>$ at large distances
is given by a bilinear combination of the beta functions of the coupling
constants taken at the scale $1/|x-y|.$
To determine the asymptotic behaviour of this correlator
it is sufficient to consider the asymptotic behaviour of the
coupling constants near the
critical values $\alpha=\alpha_*$ and $\lambda=\lambda_*$.
We assume that the beta functions have simple zeros at the critical point
i.e.
\eqn\betascr{\beta_{\alpha}=\beta_{\alpha \alpha}'  (\alpha-
\alpha_*)+\beta_{\alpha \lambda}' (\lambda-\lambda_*),}
$$\beta_{\lambda}=\beta_{\lambda \alpha}' (\alpha-\alpha_*)+
\beta_{\lambda \lambda}' (\lambda-\lambda*) ,$$
where the constants
$\beta_{\alpha \alpha}'$, $\beta_{\alpha \lambda}'$,
$\beta_{\lambda \alpha}'$
and
$\beta_{\lambda \lambda}'$ are the components of the matrix  
$\beta_{ij}'$.
An explicit computation shows that
\eqn\andimmag{<a_\mu(x)a_\nu(y)>\sim {1\over |x-y|^{2\beta_{min}'+6 }},}
where $\beta_{min}'$ is the minimal eigenvalue of the matrix
$\beta_{ij}'$.
It follows from eq. \andimmag\ that the anomalous dimension of the  
magnetic
Konishi current
is given by $\beta_{min}'$ which  is to be identified with the anomalous
dimension
$\delta$ of the Konishi current in the electric theory.
Thus we get
\eqn\univ{\delta=(\beta' (\alpha_*))_{electric}= (\beta_{min}'( \alpha_*,
\lambda_*))_{magnetic}.}

Eq. \univ\ is the main result of this section.
It is interesting  to consider the strong coupling regime
$(2N_f-3N_c)/N_c<<1$.
In this case the magnetic theory is weakly coupled, we can compute
$\beta_{min}'$
in perturbation theory \ksv , and thus we can explicitly
obtain the slope of the beta function in the strongly coupled
electric theory
\eqn\wc{\delta=(\beta ' (\alpha_*))_{electric}=
(\beta_{min}'( \alpha_*, \lambda_*) )_{magnetic}={28\over
3}\left({3\over 2}
-{N_f  \over N_c}\right)^2 .}
It is instructive to compare the above result with the behavior of  
the theory
outside the conformal  window.
The electric-magnetic duality  implies  identical
behaviour of the correlators of the Konishi currents at large  
distances in both
electric and magnetic descriptions.
Since in the magnetic description the theory for $N_f<3N_c$ is free in
the infrared, duality predicts that the same is true for the electric
formulation where the theory confines.
In turn this implies that the functions $\phi_{1,2}$ in the  
correlator \KK\
in the electric theory are  proportional to
$(\alpha- 2\pi/(N_f -N_c))$ ~at $\alpha\to \alpha_*=2\pi/(N_f -N_c)$
in order to cancel the pole in the $z\sim \beta (\alpha)$ factor of  
the Konishi
current.
Therefore the Konishi current has a canonical dimension in the infrared,
outside the conformal window.

\newsec{Quantum conformal algebra}

In this section we collect some simple observations about
CFT$_4$ that were
stimulated by the investigation carried out so far.
By using these observations we shall argue for the existence of a
duality map
between distinguished deformations (in the infinitely small
neighbourhood of
the critical point) which follow from the structure of the superconformal
algebra.
We shall see that  such a deformation is a close relative of the Konishi
current
(and perhaps coincides with the latter) and plays an important role
in CFT$_4$.
This observation may explain the electric-magnetic universality.

There are two strategies for studying conformal field theories. One is
the strategy
that we have pursued in the previous sections,  starting from the
off-critical
theories and flowing to the fixed point. We have seen how
electric-magnetic duality can be used concretely to
compute nontrivial
physical quantities that otherwise cannot be computed (in the
strong coupling regime, for example).
Another strategy to study conformal field
theories is to define them in an axiomatic-algebraic way and then
classify
the unitary
theories, the representations, etc., more or less as in two dimensions.
In this section we make some remarks about CFT$_4$
that could be the basis for the axiomatic approach to conformal
field theories
in four dimensions (CFT$_4$). These remarks could also have other
consequences in the context of electric-magnetic duality.
They give a general picture of the role of
anomalous currents at the critical point and allow us to
argue that universality is a general property of CFT$_4$.
We note that unitary representations of the classical (super) conformal
four-dimensional algebras have been studied in ref. \ref\mack{G. Mack,
Comm. Math. Phys. {\bf 55} (1977) 1\semi
M. Flato and C. Fronsdal, Lett. Math. Phys. {\bf 8} (1984) 159\semi
V.K. Dobrev and V.B. Petkova, Phys. Lett. {\bf 162 B} (1985) 127.}.

We consider the notion of quantum (super)conformal algebra.
The quantum conformal algebra
is to the classical conformal group $SO(4,2)$
as, in two
dimensions, the centrally-extended Virasoro algebra is to the
non-extended
one. In other words, we think it is useful to study
CFT$_4$ at the level of OPE's, and not simply at the level of
the classical conformal group\foot{OPE's in dimensions $2<d<4$ have been
studied recently by A. Petkou in \ref\PET{A. Petkou, Phys. Lett. {\bf B359}
(1995) 101, and
hepth/9602054.}. We thank A. Petkou for bringing these
references to our attention.}.
We recall that CFT$_4$ is a useful notion \ref\CFL{A. Cappelli, D.
Friedan and
J. Latorre,
Nucl. Phys. {\bf B 352} (1991) 616.},
namely it is possible to extract information about correlators
although the
classical conformal
algebra is not infinite dimensional as it is in two dimensions.

It turns out that:

i) the OPE of the stress-energy tensor $T_{\mu\nu}$ with itself
does not close;
another operator $\Sigma$ is brought into the algebra;

ii) there are {\sl two} central charges, $c$ and $c'$, one related to
$T_{\mu\nu}$, the other one to $\Sigma$; in the supersymmetric case
they count both the numbers of effective vector
multiplets and
matter multiplets;
$c$ is related to the conformal anomaly in an external
gravitational field
\CFL ;

iii) $\Sigma$ is expected to have an anomalous dimension, in general;
in the free supersymmetric
case, it is precisely the Konishi current; the results of the
previous sections suggest that $\Sigma$ may coincide with the
Konish current
in an interacting CFT$_4$.

We begin with the simplest example of CFT$_4$, a free
massless scalar. We have
$$
{\cal L}={1\over 2}\partial_\mu\varphi\partial_\mu\varphi,\quad\quad
<\varphi(x)\varphi(0)>={1\over x^2},$$$$
T_{\mu\nu}={2\over 3}\partial_\mu\varphi\partial_\nu\varphi-{1\over
6}\delta_{\mu\nu}
(\partial_\rho\varphi)^2-{1\over
3}\varphi\partial_\mu\partial_\nu\varphi.
$$
We have written the so-called improved stress energy tensor $T_{\mu\nu}$
\ref\coleman{C.G. Callan, Jr., S. Coleman and R. Jackiw,
Ann. Phys. (N.Y.) {\bf 59} (1970) 42.},
which
differs from the canonical one $T^{c}_{\mu\nu}$ by a term like
$\partial_\rho\chi_{\rho\mu\nu}$, with
$\chi_{\rho\mu\nu}=-\chi_{\rho\nu\mu}$.
The improved tensor is symmetric,
conserved and traceless.  Terms proportional to the field equations
${\,\lower0.9pt\vbox{\hrule \hbox{\vrule
height 0.2 cm
\hskip 0.2 cm \vrule height 0.2 cm}\hrule}\,}\varphi$ are omitted
here, since
they do not affect the arguments to be presented.
$T_{\mu\nu}$ is related to the curved space
conformally invariant lagrangian
${\cal L}=\sqrt{g}\,(1/2
\,g^{\mu\nu}\partial_\mu\varphi\partial_\nu\varphi-1/12 \,\varphi^2 R)$
\coleman .
At the level of commutators, $T_{\mu\nu}$
generates the correct conformal group $SO(4,2)$ \coleman .
At the level of OPE's, however, we note that there is a term
containing the
unity operator. This term is
associated with
a notion of central charge in four dimensions and is analogous to
the central
extension of CFT$_2$.
In addition a new operator $\Sigma$ appears, without
which the
OPE's do not close. In this case we are
dealing with $\Sigma=\varphi^2$.
Indeed, we find
\eqn\OPEuno{T_{\mu\nu}(x)\,T_{\rho\sigma}(0)=-c \,{1\over 360}\,
X_{\mu\nu\rho\sigma}\!\!\left({1\over x^4}\right)-
{1\over  36}\,\Sigma(0)\,X_{\mu\nu\rho\sigma}\!\!\left({1\over
x^{4-h}}\right)+
{\rm less\,\, singular},}
with $c=1$ and $h=2$,
where
$$X_{\mu\nu\rho\sigma}=2{\,\lower0.9pt\vbox{\hrule \hbox{\vrule
height 0.2 cm
\hskip 0.2 cm \vrule height 0.2 cm}\hrule}\,}^2\delta_{\mu\nu}
\delta_{\rho\sigma}-3{\,\lower0.9pt\vbox{\hrule \hbox{\vrule height
0.2 cm
\hskip 0.2 cm \vrule height 0.2
cm}\hrule}^2\,}\,(\delta_{\mu\rho}\delta_{\nu\sigma}+
\delta_{\nu\rho}\delta_{\mu\sigma})-2{\,\lower0.9pt\vbox{\hrule
\hbox{\vrule
height 0.2 cm \hskip 0.2 cm \vrule height 0.2
cm}\hrule}\,}\,(\partial_\mu
\partial_\rho
\delta_{\nu\sigma}+\partial_\nu \partial_\rho \delta_{\mu\sigma}+
\partial_\mu \partial_\sigma \delta_{\nu\rho}+\partial_\nu
\partial_\sigma
\delta_{\mu\rho})
$$$$-2{\,\lower0.9pt\vbox{\hrule \hbox{\vrule height 0.2 cm \hskip 0.2 cm
\vrule height 0.2 cm}\hrule}\,}\,(\delta_{\mu\nu}\partial_\rho
\partial_\sigma+
\delta_{\rho\sigma}\partial_\mu \partial_\nu)-4\partial_\mu \partial_\nu
\partial_\rho
\partial_\sigma.$$
$c$ is the first central charge that we find.
Less singular terms are easily workable.
The expression $X_{\mu\nu\rho\sigma}$ is uniquely fixed by the symmetries
in the indices and by
requirements $\partial_\mu
X_{\mu\nu\rho\sigma}=0$
and $X_{\mu\mu\rho\sigma}=0$.

To check  closure, we  examine the OPE's
of $\Sigma$. We obtain
$$
\Sigma(x)\Sigma(0)={2c'\over x^{2h}}+{2\over
x^h}\Sigma(0)+\cdots$$\eqn\OPEdue{
T_{\mu\nu}(x)\Sigma(0)=-{h\over
3}\,\Sigma(0)\,\partial_\mu\partial_\nu\!\!\left({1\over x^2}\right)
+\cdots}
In the case at hand $c'=c$, but this is not true in general. $c'$
is a second central charge.
The case of $n$ free massless scalars $\varphi^i$, $i=1,\ldots n$,
is straightforward: $c=c'=n$, $h=2$ and $\Sigma=\varphi^i\varphi^i$.
This is sufficient to show that $c$ and $h$ are independent parameters
of CFT$_4$.

Schematically, we can expect, for bosonic theories, something like
$$
T(x) T(0)={c\over x^8}+{\Sigma\over x^{8-h}}+{\partial \Sigma\over
x^{7-h}}+
{T\over x^4}+\cdots,$$$$
T(x)\Sigma(0)={h\Sigma\over x^4}+{\partial\Sigma\over
x^3}+{\partial^2\Sigma\over
x^2}+
{T\over x^h}+\cdots,$$
\eqn\qcft{\Sigma(x)\Sigma(0)={c'\over x^{2h}}+{\Sigma(0)\over
x^h}+{\partial\Sigma\over
x^{h-1}}}
or appropriate superextensions.
\OPEuno\ and \OPEdue,
or \qcft, or their superextensions, define what we call the {\sl
quantum conformal algebra}.

In the complex case, $\Sigma$ becomes $\bar\varphi\varphi$ and in the
supersymmetric case, it becomes the Konishi current.
In the free theory, the Konishi current
does not depend on the vector multiplets.
We shall see that in the absence of matter multiplets the
supercurrent  $J_{\alpha\dot\alpha}=W_\alpha
\bar{W}_{\dot\alpha}$
is sufficient for closure; no additional operator appears for vector
multiplets.
Thus  $c'$ is sensitive to matter only, $c$ is sensitive to
both matter and gauge multiplets .
We believe that this is an important property of CFT$_4$, because
it allows us
to count separately the matter multiplets and the vector multiplets.

The results of the previous sections suggest
that  the non-closure of the OPE
of $T_{\mu\nu}$ with itself is a general property of CFT$_4$,
and that  in general, the OPE's are like \OPEuno\ and \OPEdue\
with $c$, $c'$ and $h$ generic.
Physical states and primary fields have to be defined
with respect to the complete algebra, of course.
Defining a primary scalar field $\phi$ of conformal dimension $\Delta$ by
\eqn\primary{
T_{\mu\nu}(x)\phi(0)=-{\Delta\over
3}\,\phi(0)\,\partial_\mu\partial_\nu\!\!\left({1\over x^2}\right)+{\rm
less\,\, singular},}
would imply that the primary fields of a free
scalar in CFT$_4$ are $\varphi^k$, with weight $k$,
to be compared with $\partial\varphi$ and ${\rm e}^{\alpha \varphi}$ in
CFT$_2$. However, due to the nonclosure of $T_{\mu\nu}$ with itself,
this is not the full story. With respect to $\Sigma$,
$\varphi^{2k+1}$ mixes with
$\varphi^{2k-1}$.
Thus it is not clear what   the proper notion of primary field is.
Perhaps  one has to
consider primary {\sl sets} of fields, such as
$\{\varphi^{2k+1},\varphi^{2k-1},\ldots,\varphi\}$.

For the case of free
fermions we have
$$
{\cal L}={1\over 2}(\bar
\psi\partial\!\!\!\slash\psi-\partial_\mu\bar\psi
\gamma_\mu\psi),\quad\quad
<\psi(x)\bar\psi(0)>={x\!\!\!\slash \over x^4},$$$$
T_{\mu\nu}={1\over 2}(\bar \psi\gamma_{\mu}\partial_\nu\psi+\bar
\psi\gamma_{\nu}\partial_\mu\psi-\partial_\mu\bar\psi\gamma_\nu\psi-
\partial_\nu\bar\psi\gamma_\mu\psi).
$$
$x\!\!\!\slash=x^\mu\gamma_\mu$.
The analogue of \OPEuno\ exhibits a central term
with $c=6$, so that $c={3\over 2}$ for any real on-shell component.
The $\Sigma$-term involves the axial current
$J^5_\mu=\bar\psi\gamma_5\gamma_\mu\psi$.
We find
\eqn\TTferm{
T_{\mu\nu}(x)T_{\rho\sigma}(0)=-c \,{1\over 360}\,\,
X_{\mu\nu\rho\sigma}\!\!\left({1\over x^4}\right)+\sim
\varepsilon_{\mu\rho\alpha\beta}J_5^\beta  
(0)\,\partial_\nu\partial_\sigma
\partial_\alpha\left({1\over x^2}\right)+\ldots}
For the case of  vectors we have
$$
T_{\mu\nu}=F_{\mu\rho}F_{\rho\nu}+{1\over
4}\delta_{\mu\nu}F^2+s-{\rm exact \,\,terms}.
$$
One finds $c=12$. The $s$-exact terms do not affect $c$ and only
introduce
additional $s$-exact operators in the OPE's.
There is no $\Sigma$-term and closure is achieved.
As we shall see below, analogous conclusions hold for the
supersymmetric vector
multiplet, where the
single superfield  $J_{\alpha\dot\alpha}=W_\alpha W_{\dot\alpha}$
is still sufficient for closure and so again $\Sigma=0$ and $c'=0$.
It is also meaningful to speak about primary fields, in this case.
For example,
$W_\alpha$ is a primary field with weight 3/2.
It is not clear  whether the free vector multiplet is the only CFT$_4$
with $\Sigma=0$ and $c'=0$.
Theories with $c'=0$ would appear effectively as pure Yang-Mills
conformal
field theories.
The other possible case is a topological theory, where
$c$ and $c'$ both vanish.

In a generic nonsupersymmetric conformal field
theory containing scalars, fermions and
vectors the central charges are three. They allow to count both the  
scalar,
fermion and vector effective degrees of freedom separately,
via the operators $T_{\mu\nu}$, $\varphi^2$ and $J_5^\mu$.
Supersymmetry reduces the number
of independent central charges to two, so that
for an N=1 free theory with $m$ matter multiplets and $v$ vector
multiplets, we have $c=15 v+ 5m$, $c'=5m$, which encode
the effective numbers of  matter and vector multiplets.

In free N=4 supersymmetric Yang-Mills theory,  $c=30n$, $c'=15n$,  
$n=v=m/3$.
Moreover, since $c$ is related to
an anomaly\foot{We stress here that N=4 supersymmetric
Yang-Mills theory
is indeed  anomalous in an external background gravitational field.
The anomaly vanishes for Ricci-flat  backgrounds, but not in general.
The importance of this fact is quite evident,   since it is
reflected in the non-vanishing of $c$.}
\ref\FB{See, for example, F. Bastianelli, {\it Tests for C-theorems in 4D},
Phys. Lett.  {\bf B369} (1996) 249,
hep-th/9511065, and references therein.}
(which obviously does not have
any multiloop corrections in this theory),  $c=30n$ also
in the interacting case, at least at the
perturbative level (there could be instanton corrections).
It is not  clear whether $c'$ is also unaffected by turning on the
interaction.
S-duality tells us that $c'=15n$ also in the infinitely strong
coupling regime.
Note that a deformation
$\lambda\Sigma$
cannot preserve N=4
supersymmetry, since
the only candidate in that case would be $\Sigma={\cal L}$, which is
not the case in the free theory.
Instead, general considerations show that
$\Sigma$ is proportional to
the Konishi superfield also in the weakly coupled theory.
It would be interesting to compute the corresponding $c'$.
Moreover, $\Sigma$ has an anomalous dimension
$\beta'_{\lambda}\neq 0$.
Indeed in the weak coupling limit an explicit computation
gives (for $SU(N_c)$ gauge
group) $\beta'_{\lambda}=3\alpha N_c/\pi$.
In general due to $S$ duality the parameters $\beta'_{\lambda}$ and
$c'$ must
be
real non-singular functions of $J(q)$ and $J(\bar{q})$,
where $J(q)$ is a generator of modular functions of
$q=\exp (2\pi i\tau)$, $\tau=\theta/2\pi +4\pi i/g^2$;  $\theta$ is
a theta angle in front of $F\tilde{F}$ in the Lagrangian (see, for
example, \ref\vw{C. Vafa and E. Witten, Nucl. Phys. {\bf B 431}  
(1994) 3.}).
A dependence of $c'$ on the coupling constant of N=4 theory would  
imply that
the effective numbers of chiral and vector multiplets (in the N=1 sense)
changes with $g$ while the total number does not.
Further analysis of marginal deformations of SQFT$_4$ can be found
in \ref\AFGJ{D. Anselmi, D.Z. Freedman, M.T. Grisaru and A. Johansen,
{\it Universality of the operator product expansions of SCFT$_4$},
preprint HUTP-96-A037; hep-th/9608125.}.

In conclusion, while the quantum conformal algebra  can provide us,
through the identification of two CFT$_4$'s with a map between
the two operators $\Sigma_{el}$ and $\Sigma_m$, the
considerations of the
previous sections show how to map the Konishi currents.
It is reasonable to expect that in general $\Sigma$ is related to or even
coincides with
the Konishi operator.
At any rate the understanding of this relation should  be a source of
additional insight.
The maps of anomalous dimensioned operators are
also maps of deformations of the conformal theories.
The  appearance of one
anomalous operator in the OPE of the stress-energy tensor helps to
clarify the role of anomalous currents
in CFT$_4$.

For completeness we  give the corresponding results in superspace,
examining
the OPE's of the supercurrent and Konishi current for a free chiral
scalar
superfield (scalar
multiplet), and for a  gauge real scalar superfield (vector
multiplet). We use
the conventions of
{\it Superspace}  \ref\Superspace{S.J. Gates, M.T. Grisaru, M.
Ro\v{c}ek and W.
Siegel,
{\it Superspace},  Benjamin-Cummings, Reading MA, 1983.}.
For a chiral scalar superfield with action
$\int d^4x d^4 \theta \bar{\Phi} \Phi$ the propagator, supercurrent, and
Konishi current
are, respectively, with $z \equiv (x, \theta , \bar{\theta})$,
\eqn\props{
 \eqalign{<\Phi (z) \bar{\Phi} (z') > &= -\bar{D}^2 D^2 {1 \over
(x-x')^2}
\delta^4 (\theta - \theta ') =
 -{1 \over s^2} \cr
 <\bar{\Phi} (z) \Phi (z') >
&= -D^2 \bar{D}^2 {1\over(x-x')^2} \delta^4 (\theta - \theta ') =
 -{1 \over \bar{s}^2}  \cr
 J_{\alpha \dot{\alpha}}
&= {1\over 6} [D_\alpha  , \bar{D}_{\dot{\alpha}}] \Phi \bar{\Phi}
-{1 \over 2}
\Phi i \buildrel \leftrightarrow \over \partial_{\alpha \dot{\alpha}}
\bar{\Phi}
\cr
J&= \Phi \bar{\Phi}\cr}
}
while for the vector multiplet with  field strength $W_\alpha =i
\bar{D}^2D_\alpha V$ and  action $\int d^4x d^2 \theta W^2 + h.c.$,
 quantized
in Feynman gauge, the propagator and supercurrent are
\eqn\propv{
\eqalign{<V(z) V(z')>&=  {1 \over (x-x')^2} \delta^4 (\theta -
\theta')  \cr
 J_{\alpha \dot{\alpha}}& =  W_\alpha\bar{W}_{\dot{\alpha}}\cr} .
}
Here  $(x-x')^2 = {1 \over 2} (x-x')^{\alpha \dot{\alpha}} (x-x')_{\alpha
\dot{\alpha}}$, while
 $s^2 = {1 \over 2} s^{\alpha \dot{\alpha}} s_{\alpha \dot{\alpha}}$,
$\bar{s}^2 = {1 \over 2} \bar{s}^{\alpha \dot{\alpha}} \bar{s}_{\alpha
\dot{\alpha}}$,
 where the chiral and antichiral superspace intervals
are given by
$$
\eqalign{ s_{\alpha \dot{\alpha}} &= (x-x')_{\alpha \dot{\alpha}}
+{i \over
2}[ \theta_\alpha (\bar{\theta} - \bar{\theta}' )_{\dot{\alpha}} +
\bar{\theta}'_{\dot{\alpha} } (\theta - \theta ')_\alpha ] \cr
\bar{s}_{\alpha \dot{\alpha}} &= (x-x')_{\alpha \dot{\alpha}}  +{i
\over 2}[
\bar{\theta}_{\dot{\alpha} }({\theta} - {\theta}' )_{\alpha} +
{\theta}'_\alpha (\bar{\theta} -\bar{ \theta} ')_{\dot{\alpha} }]\cr}
$$
We note the relations
$$
\eqalign{D_\beta s^{\alpha \dot{\alpha}} &= i \delta_\beta^{~\alpha}
(\bar{\theta} - \bar{\theta}')^{\dot{\alpha}}
{}~~~~,~~~~\bar{D}_{\dot{\beta}} s^{\alpha \dot{\alpha}} =0\cr
\bar{D}_{\dot{\beta} } \bar{s}^{\alpha \dot{\alpha}} &= i
\delta_{\dot{\beta}}^{~\dot{\alpha}} ({\theta} - {\theta}')^\alpha
{}~~~~,~~~~ D_\beta \bar{s}^{\alpha \dot{\alpha}} =0\cr}
$$
Note also the relations
$$
D^2 {1 \over s^2} = \delta^{(4)}(x-x') \delta^{(2)}
(\bar{\theta}-\bar{\theta}') ~~~~,~~~~\bar{D}^2 {1\over \bar{s}^2}=
\delta^{(4)}(x-x') \delta^{(2)}(\theta- \theta ' )
$$
and, up to contact terms,
$$
< \bar{D}_{\dot{\alpha}} \bar{\Phi} D_{\alpha} \Phi > = -i
\partial_{\alpha
\dot{\alpha}}
{1 \over s^2} .
$$
The Taylor expansions are
$$
\eqalign{\Phi (z') &=\Phi (z) +s^{\alpha \dot{\alpha}} \partial_{\alpha
\dot{\alpha}} \Phi (z) + (\theta - \theta ')^\alpha D_\alpha \Phi
(z) + \cdots
\cr
\bar{\Phi}(z') &=\bar{\Phi} (z) +\bar{s}^{\alpha \dot{\alpha}} \partial_{\alpha
\dot{\alpha}} \bar{\Phi} (z) + (\bar{\theta} -
\bar{\theta} ')^{\dot{\alpha}}\bar{D}_{\dot{\alpha}} \bar{\Phi} (z) +
\cdots\cr}  .
$$

We determine now the OPE's setting, for convenience, $z' =0$. For
the Konishi
current, performing double and single contractions
of the chiral superfields, we obtain
\eqn\JJ{
J(z) J(0)  ={1\over s^2 \bar{s}^2} -J(0) \left({1 \over s^2} +{1 \over
\bar{s}^2} \right)
+\cdots
}
For the OPE of the supercurrent with the Konishi current we obtain
\eqn\KJ{
J_{\alpha \dot{\alpha}} (z) J(0) =
- {J(0)
\over 3}
\partial_{\alpha \dot{\alpha}}\left({1\over s^2}-{1 \over
\bar{s}^2}\right)
+\cdots
}

Finally, for the OPE of two supercurrents we have
\eqn\JJmat{
J_{\alpha \dot{\alpha}}(z) J_{\beta \dot{\beta}}(0) =
-{1\over 3}{ s_{\alpha\dot{\beta}}  \bar{s}_{\beta \dot{\alpha}}
\over (s^2
\bar{s}^2)^2}
+{J(0)\over 9}
\partial_{\alpha \dot{\alpha}}\partial_{\beta
\dot{\beta}}\left({1\over s^2}
+{1\over \bar{s}^2 }\right) + \cdots
}
We note the appearance of the Konishi current in the OPE of two
supercurrents.
Less singular terms involve derivatives of  the Konishi current and the
supercurrent.

For the vector multiplet the OPE of two supercurrents is simply
\eqn\JJv{
J_{\alpha \dot{\alpha}}(z) J_{\beta \dot{\beta}}(0) = -{ s_{\alpha
\dot{\beta}}
\bar{s}_{\beta \dot{\alpha}}\over (s^2 \bar{s}^2)^2}
-J_{\alpha \dot{\beta}}(0) {i \bar{s}_{\beta \dot{\alpha}}\over
(\bar{s}^2)^2}
+J_{\beta \dot{\alpha}}(0) {i s_{\alpha \dot{\beta}}\over (s^2)^2}  
+\cdots
}
Note that no term similar to the Konishi term in \JJmat \ appears here.
It is straightforward to check  that the above expressions vanish
when, e.g.,
$D^\alpha$
acts on them.

\par
\newsec{Conclusions}

In this concluding section
we recapitulate the situation and make some further comments.

In this paper we have  studied properties of the (anomalous)
Konishi current
in N=1  SQCD, away from the critical point,  and constructed the
duality map for the Konishi supermultiplet.
We have found that contrary to the duality map constructed previously for
the primary chiral ring the present map
involves a consideration of  the space of  off-critical deformations.
We have related these observations to general properties of CFT$_4$.

Away from the fixed point the Konishi current is the only renormalizable
deformation of electric conformal SQCD.
Indeed such a deformation can be reexpressed as a change of the
gauge coupling constant due to the Konishi anomaly.
Thus the duality map of the Konishi operator of the electric theory
implies that a particular renormalizable deformation of the
electric theory corresponds
to a renormalizable deformation of the magnetic one.
This means that in the {\it infinitely small} neighbohood of a  
critical point
there
exists a map from the electric off-critical theory to the magnetic one.
It is worth emphasizing that such a map respects the
renormalizability (up to less relevant operators), thus demonstrating a
universality of the subleading (non-critical)
behaviour of the theory near critical points.
This  universality manifests itself in the matching of the slopes
of the beta functions in the electric and magnetic theories as exhibited
in eq. \univ .

At the level of OPE's,
we have found that an operator $\Sigma$, equal to the Konishi
current at least in free supersymmetric theories,
mixes with the stress-energy tensor.
We believe that quite generally, its existence is related to the
presence of anomalous currents in  CFT$_4$.
We have noted the role of  $\Sigma$ in a distinction  between
matter multiplets and vector multiplets.
Finally, in the context of electric-magnetic duality, we note that
the matching of $\Sigma_{el}$ and $\Sigma_{mag}$ is also
a correspondence between off-critical deformations.
We believe that these observations may be useful in an
axiomatic-algebraic approach to
four-dimensional conformal field theories.

The above observed universality suggests that perhaps one can construct
an electric-magnetic duality map even  far away  from criticality.
{\it A priori} it is clear that such a map  must be very
complicated since one cannot in general
avoid a non-renormalizable extension of the theory at finite  
distances away
from the fixed point.
Indeed, the renormalizable asymptotically free electric and magnetic
theories cannot be equivalent at an arbitrary scale since, for  
example, there
is no (at least naively) equivalence  at the ultraviolet fixed point.
On the other hand it is plausible to expect
the existence of a more general map which includes
non-renormalizable deformations of the theory.
But renormalizability of the theory is a very important constraint  
in four
dimensions since in general there is no way to define a
non-renormalizable theory.
Therefore, we must understand a non-renormalizable deformation of
the theory as the effect of heavy particles which may become  
relevant in some
regime of the low-energy theory.
An explicit introduction of such  heavy fields can transform the
deformed theory into a renormalizable form.

\newsec{Acknowledgements}
The authors are grateful to M. Bershadsky,  C. Vafa,  A. Vainshtein and
D. Zanon for interesting and valuable discussions, D. Kutasov for useful
correspondence and D. Freedman for reading the manuscript and for
comments.
D.A. and A.J. are partially supported by the
Packard Foundation and by NSF grant PHY-92-18167.
The work of A.J. is also supported in part by NATO grant GRG 930395.
The work of M.G. is partially supported by NSF grant PHY-92-22318.

\newsec{Appendix}

We outline here the perturbative checks one can perform on the
results of section 3.
The leading contribution proportional to $\log \Lambda$
to the correlator  $<a_{\mu}(x) a_{\nu}(y)>$ is due to the diagram  
in Fig. 1.

\vskip .2in
\let\picnaturalsize=N
\def\picsize{2.5in}
\def\picfilename{crfig1.eps}
\ifx\nopictures Y\else{\ifx\epsfloaded Y\else\fi
\global\let\epsfloaded=Y
\centerline{\ifx\picnaturalsize N\epsfxsize \picsize\fi
\epsfbox{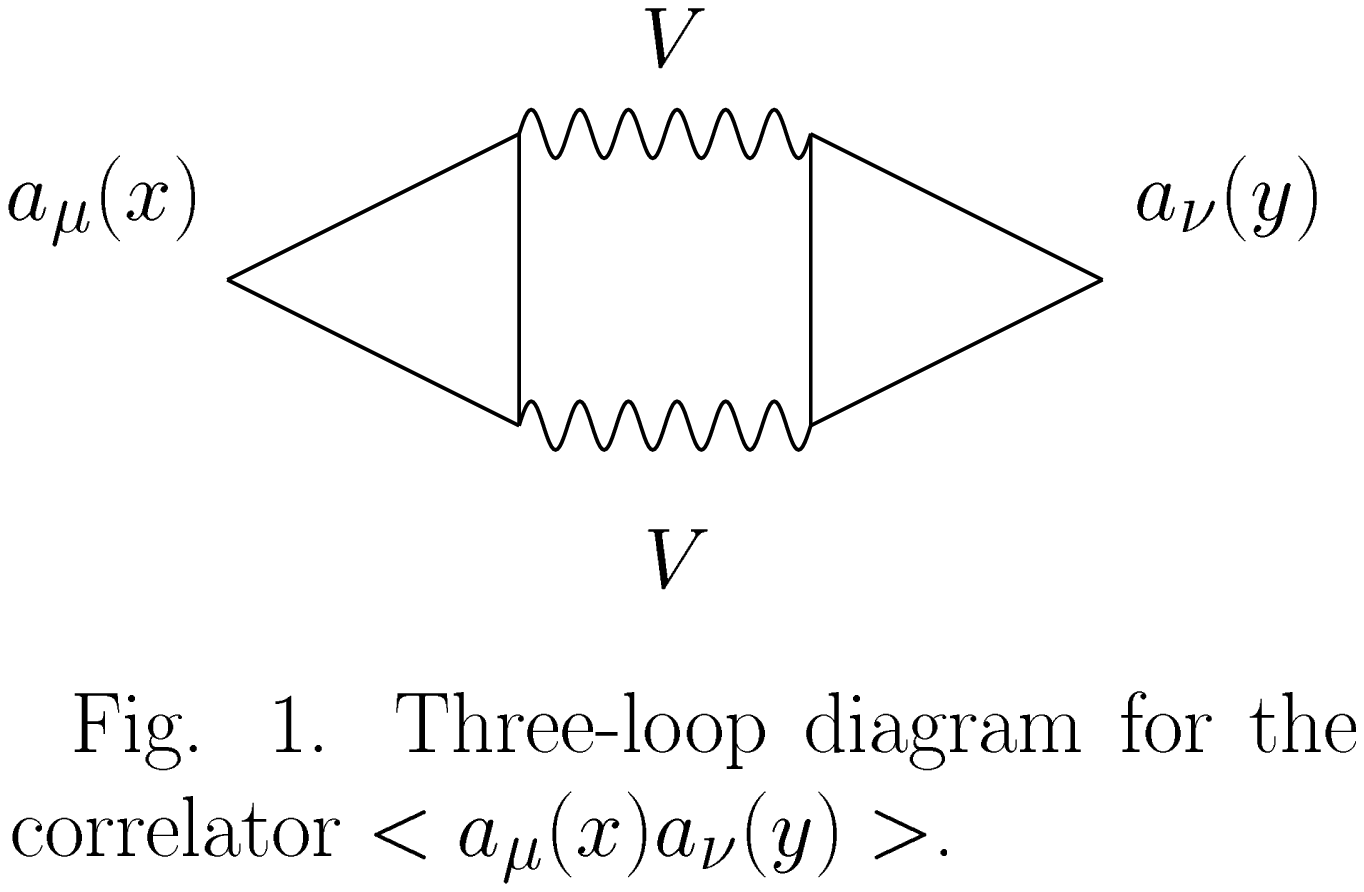}}}\fi

In order to extract this contribution we have to consider two
asymptotic domains in the integral corresponding to the diagram in
Fig. 2.
\vskip .2in

\let\picnaturalsize=N
\def\picsize{4.0in}
\def\picfilename{crfig2.eps}
\ifx\nopictures Y\else{\ifx\epsfloaded Y\else\fi
\global\let\epsfloaded=Y
\centerline{\ifx\picnaturalsize N\epsfxsize \picsize\fi
\epsfbox{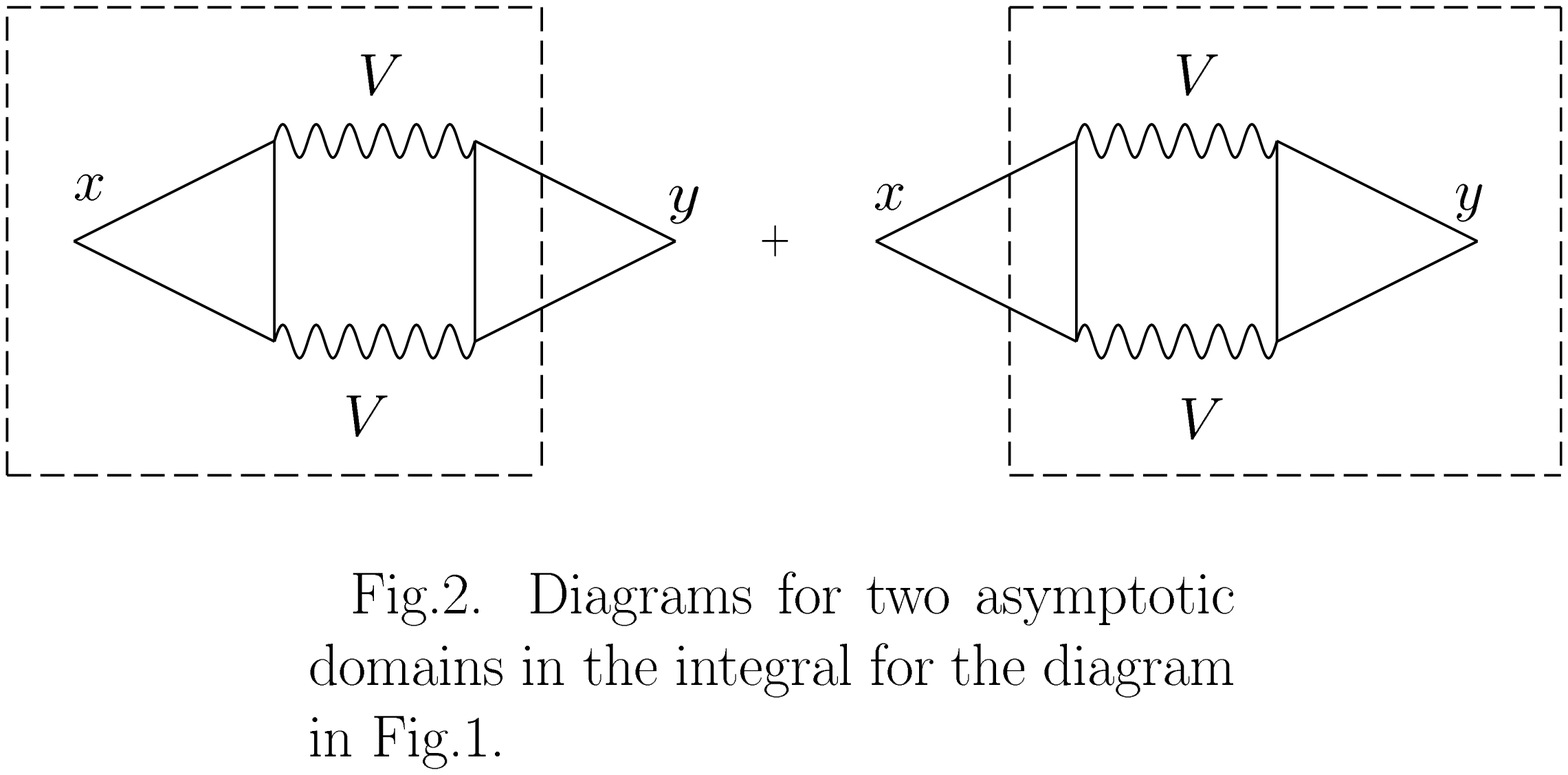}}}\fi

The short distance contribution of each of the marked subdiagrams
in Fig. 2 corresponds to
the leading contribution to the anomalous dimension of the current $J$.
Therefore these subdiagrams can be effectively shrunk to
point-like vertices proportional to
$z-1=b_2\alpha^2\log \Lambda|x-y|$ as shown in Fig. 3,
where  $b_2$ is the  two-loop coefficient of
the beta function.
Taking into account the tree level diagram we get
in this approximation $1 +2b_2 \alpha^2\log \Lambda|x-y|$ for the
correlator
$<a_{\mu}(x) a_{\nu}(y)> $.
By using the renormalization group we restore the multi-loop
expression for the correlator.
(Note however that this computation is formally justified only
at $g^2\log|x-y|\leq 1$.)
Renormalizing it with the factor $(\beta
(\alpha_0)/\beta_1(\alpha_0))^2$ we get eq. \KK \
with $\phi _{1,2}(\alpha)=1$.
The triviality of the function $\phi (\alpha)$ can be of
course due to our approximation.

\vskip .2in
\let\picnaturalsize=N
\def\picsize{3.5in}
\def\picfilename{crfig3.eps}
\ifx\nopictures Y\else{\ifx\epsfloaded Y\else \fi
\global\let\epsfloaded=Y
\centerline{\ifx\picnaturalsize N\epsfxsize \picsize\fi
\epsfbox{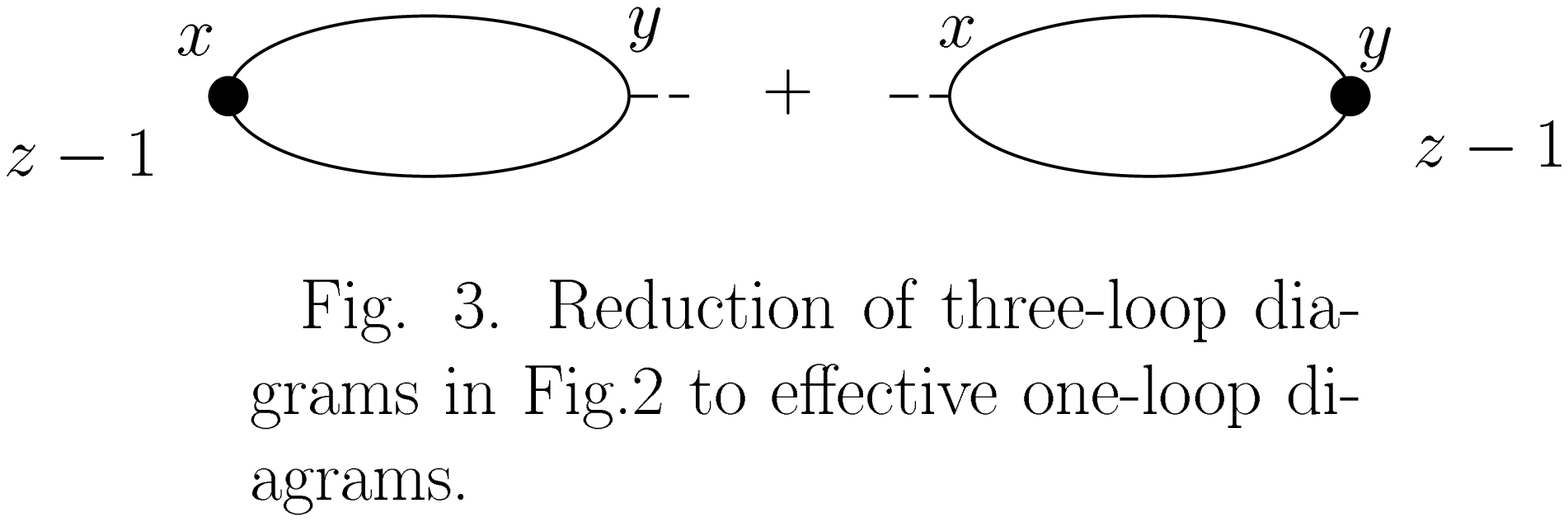}}}\fi

Similarly one can compute the correlators of other currents.
In the conformal theory limit we are interested in the operators
which are
invariant under
the renormalization group.
Therefore we consider the conserved $\tilde{R}_{\mu}$ current, which is a
linear
combination \ksv \
of the current $R_{\mu}$ which enters the $J_{\alpha\dot{\alpha}}$
superfield
and the Konishi axial current $a_{\mu}$
$$\tilde{R}_{\mu} =R_{\mu}+\left(1-{3N_c\over
N_f}-\gamma\right)a_{\mu}.$$
It is easy to check by an explicit computation of the correlator
$<\tilde{R}_{\mu}(x) \tilde{R}_{\nu}(y)>$  that in
perturbation theory
the anomalous contributions $\sim 1/ |x-y|^{6+\beta'(\alpha_*)}$ and
$\sim 1/ |x-y|^{6+2\beta'(\alpha_*)}$
are cancelled, and hence the correlator is transverse
$$ <\tilde{R}_{\mu}(x)\tilde{R}_{\nu}(y)>\sim
(g_{\mu\nu}\partial^2-\partial_{\mu}\partial_{\nu})
{1\over |x-y|^4}.$$
This is as expected since  the conserved current $\tilde{R}_{\mu}$  
should have
canonical conformal dimension; there is no $z$ factor for the
correlator in question.

The scaling dimension of other components
of the operators $J$ and ${\rm Tr} \;W^2$ can be easily determined by
supersymmetry.
Note that the scaling of ${\rm Tr}\; W^2$ agrees with the fact that this
operator
is not a primary chiral one.
This also follows from the anomaly equation for the Konishi
supermultiplet.

\listrefs

\end